\documentclass[12pt,a4paper]{article}
\renewcommand{\theequation}{\arabic{section}.\arabic{equation}}

\usepackage{amsmath,amssymb}
\usepackage{graphicx,color}
\usepackage{bbm}
\usepackage{cite}

\newcommand{\nn}{\nonumber}
\newcommand{\Eqn}[1]{&\hspace{-0.5em}#1\hspace{-0.5em}&}
\newcommand{\eqb}{\begin{eqnarray}}
\newcommand{\eqe}{\end{eqnarray}}
\renewcommand{\thefootnote}{\fnsymbol{footnote}}
\newcounter{aff}

\newcommand{\alg}[1]{\mathfrak{#1}}
\newcommand{\grp}[1]{\mathrm{#1}}

\def\comma      { \, , }
\def\period     { \, . }
\def\del        {  \partial  }
\def\bra{\langle}
\def\ket{\rangle}
\def\calO   {{\cal O}}
\def\calF   {{\cal F}}
\newcommand{\bbZ}{{\mathbb Z}}
\DeclareMathOperator{\im}{Im}

\DeclareMathOperator{\dilog}{Li_{2}}
\def\blambda{\boldsymbol{\lambda}}
\def\bLambda{\boldsymbol{\Lambda}}
\def\bmu{\boldsymbol{\mu}}
\def\bomega{\boldsymbol{\omega}}
\def\brho{\boldsymbol{\rho}}
\newcommand\bzero{\boldsymbol{0}}
\newcommand{\One}{{\bf 1}}
\def\tiln { \tilde{n} }
\def\cO{{\mathcal O}}
\DeclareMathOperator{\Li}{Li}

\def\({\left(}
\def\){\right)}
\makeatletter \@addtoreset{equation}{section} \makeatother
\renewcommand{\theequation}{\arabic{section}.\arabic{equation}}
\makeatletter
\renewcommand{\appendix}{
\renewcommand{\thesection}{\Alph{section} \hspace{-.33em}}
\renewcommand{\thesubsection}
{\Alph{section}.\arabic{subsection} \hspace{-.33em}}
\@addtoreset{equation}{subsection}
\renewcommand{\theequation}{\Alph{section}.\arabic{equation}}
\setcounter{section}{0}}
\makeatother

\begin{document}
\baselineskip=0.7cm
\begin{titlepage}

\hfill\hfill
\begin{minipage}{1.2in}
DESY 12-197 \par\noindent 
 TIT/HEP-623  \par\noindent 
 UTHEP-652  
\end{minipage}

\vspace{0.7cm}
\begin{center}
{\Large \bf\mathversion{bold} 
Null-polygonal minimal surfaces in AdS$_4$ from perturbed $W$ minimal models
}
\lineskip .6em
\vskip1.2cm
{\large Yasuyuki Hatsuda\footnote[1]{\tt  yasuyuki.hatsuda@desy.de},
           Katsushi Ito\footnote[2]{\tt ito@th.phys.titech.ac.jp} and
           Yuji Satoh\footnote[3]{\tt ysatoh@het.ph.tsukuba.ac.jp}
            }
\vskip 2em
${}^{\ast}${\normalsize\it DESY Theory Group, DESY Hamburg \\
 D-22603 Hamburg, Germany} \vskip 1 em
${}^{\ast,\dag}${\normalsize\it Department of Physics, Tokyo Institute of Technology\\
Tokyo, 152-8551, Japan} \vskip 1 em
${}^{\ddag}${\normalsize\it Institute of Physics, University of Tsukuba\\
Ibaraki, 305-8571, Japan} 
\vskip 1.5em
\end{center}
\vskip 3ex

\begin{abstract}
\vskip 2ex
\baselineskip=3.5ex

We study the null-polygonal minimal surfaces in AdS$_4$, which
 correspond to the gluon scattering amplitudes/Wilson loops in ${\cal
 N}=4$ super Yang-Mills theory at strong coupling. 
 The area of the minimal surfaces with $n$ cusps is characterized by the
 thermodynamic Bethe ansatz (TBA) integral equations or
 the Y-system of the homogeneous sine-Gordon model, 
 which is regarded as the $\grp{SU}(n-4)_4/\grp{U}(1)^{n-5}$ generalized 
 parafermion theory perturbed by the weight-zero adjoint operators.
 Based on the relation to the TBA systems of the perturbed $W$ minimal models,
 we solve the TBA equations by using the conformal perturbation theory, 
and obtain the analytic expansion of the remainder function around 
the UV/regular-polygonal limit for $n=6$ and $7$. 
We compare the rescaled remainder function for $n=6$  with 
the two-loop one, to observe that they are close to each other similarly 
to the AdS$_3$ case.
\end{abstract}

\vspace*{\fill}
\noindent
November  2012

\end{titlepage}
\renewcommand{\thefootnote}{\arabic{footnote}}
\setcounter{footnote}{0}
\setcounter{section}{0}

\baselineskip = 3.5ex

\section{Introduction}

 The AdS/CFT correspondence shows that minimal surfaces in AdS space-time 
are dual to the Wilson loops 
along their boundary  \cite{Rey:1998ik,Maldacena:1998im},  where
the area corresponds to the expectation value of the Wilson loops at strong coupling.
When the boundary is null-polygonal/light-like, 
the minimal surfaces also give the gluon scattering amplitudes
of ${\cal N}=4$ super Yang-Mills theory \cite{Alday:2007hr}, 
implying the duality between the amplitudes and the Wilson loops 
\cite{Alday:2007hr,Drummond:2007aua,Brandhuber:2007yx} 
and hence 
the dual conformal symmetry \cite{Alday:2007hr,Drummond:2006rz,Drummond:2007au} .
This dual conformal symmetry completely fixes 
the $n$-point amplitudes/Wilson loops with $n$ cusps up to $n=5$.
For $n \geq 6$, however, it allows deviation 
from the Bern-Dixon-Smirnov (BDS) formula
\cite{Bern:2005iz} by the remainder function 
\cite{Alday:2007he,Bern:2008ap,Drummond:2008aq}, 
which is a function of the cross-ratios of the cusp coordinates on the boundary.

At strong coupling, the corresponding area of the minimal surfaces 
is evaluated with the help of integrability \cite{Alday:2009yn}.
More concretely, one first solves a set of integral equations 
of the thermodynamic Bethe ansatz (TBA) form, or an associated 
Y-/T-system \cite{Alday:2009dv, Alday:2010vh, Hatsuda:2010cc}.
The cross-ratios are then expressed by its solution, i.e., the Y- or T-functions,
and consequently the main part of the remainder function is given 
by these Y-/T-functions as well as the free energy associated with the TBA system.

In a previous paper \cite{Hatsuda:2010cc}, Sakai and the present authors 
pointed out that the TBA equations for the minimal surfaces with $2\tiln$ cusps 
in AdS$_3$  coincide with those of the $\grp{SU}(\tiln-2)_2/\grp{U}(1)^{\tiln-3}$
homogeneous sine-Gordon (HSG) model \cite{FernandezPousa:1996hi} 
with purely imaginary resonance parameters. Similarly,  it was inferred there that  
the TBA equations for the minimal surfaces with $n$ cusps in AdS$_4$ 
are those of the HSG model associated with
$\grp{SU}(n-4)_4/\grp{U}(1)^{n-5}$.

These observations allow us to solve the TBA systems around 
the UV/high-temperature limit, where the two-dimensional integrable (HSG) model
reduces to a conformal field theory (CFT). 
The deviation from the UV limit  then corresponds to an integrable 
relevant/mass perturbation of the CFT. The corrections
to observables are also regarded as 
finite size effects of the two-dimensional system, which can be
computed by using the conformal perturbation theory (CPT).
By the standard procedure \cite{Zamolodchikov:1989cf}, 
one can indeed derive an analytic expansion of the free energy around the UV limit.
The Y-/T-functions are also  expanded by the CPT with boundaries 
\cite{Dorey:1999cj,Dorey:2005ak}, based on  
the relation to the $g$-function (boundary entropy)
 \cite{Affleck:1991tk}. Since the Wilson loops become regular polygonal 
in the UV limit, those expansions give an analytic expansion of the remainder 
function around this regular-polygonal limit.
For the analysis in the opposite IR/large-mass regime, 
see \cite{Alday:2009yn,Alday:2009dv,Yang:2010az,
Hatsuda:2010vr,Yang:2010as,Bartels:2010ej,Bartels:2012gq}.
 
We have carried out the above program for the minimal surfaces
embedded in AdS$_3$ \cite{Hatsuda:2011ke,Hatsuda:2011jn}. In this case, 
the relevant CFT is the  $\grp{SU}(\tiln-2)_2/\grp{U}(1)^{\tiln-3}$
generalized parafermion theory \cite{Gepner:1987sm} and,
by turning off some mass parameters so as to leave
only one mass scale (single-mass case), the TBA system is reduced to simpler ones for 
the perturbed $\grp{SU}(2)$ diagonal coset and  
minimal models. This is a key step which enables us 
to find precise values of the expansion coefficients in terms of the mass
parameters in the TBA system, through the relation 
to the coupling of the relevant perturbation (mass-coupling relation)
and the correlation functions. 
We then derived the expansion of the 8- and 10-point remainder functions
in \cite{Hatsuda:2011ke}, and that of the general $2\tiln$-point remainder function
in \cite{Hatsuda:2011jn}. 
We observed that the appropriately rescaled remainder functions are
close to those evaluated at two loops 
\cite{Brandhuber:2009da,DelDuca:2010zp,Heslop:2010kq}.

The purpose of the present work is to study the analytic expression 
of the regularized area of the null-polygonal minimal surfaces  in AdS$_4$
by extending the above results in the AdS$_3$ case.
In particular,  we derive the
 analytic expansion of the remainder function  around the UV limit by using 
the underlying integrable models and the CPT. In this case,
the corresponding TBA or  Y-system is obtained by a projection from 
that for the minimal surfaces in AdS$_5$ \cite{Alday:2010vh}.
The relevant CFT in the UV limit for the $n$-cusp surfaces is now
the $\grp{SU}(n-4)_4/\grp{U}(1)^{n-5}$ generalized parafermion theory.
The TBA systems with only one mass parameter are given by  those for  
the perturbed unitary 
$\grp{SU}(4)$ diagonal coset models and  $W$ minimal models.  
We also argue that a similar correspondence to the perturbed non-unitary diagonal coset
and $W$ minimal models holds for the systems with a pair of equal mass parameters.
These generalize the reduction in the AdS$_3$ case, and are used
to find  the precise expansion coefficients. 
Explicitly, we work out  the leading-order expansion for $n=6$ and $7$. 
In these cases, the input from the perturbed
$W$ minimal models completely determines the leading-order expansion.
For $n=6$, we also compare the rescaled remainder function 
with the two-loop one which is read off from
\cite{Anastasiou:2009kna,DelDuca:2009au,DelDuca:2010zg,Goncharov:2010jf},
to observe that they are close to each other.

This paper is organized as follows:
In section 2, we review the remainder function corresponding to 
the minimal surfaces in AdS$_4$, and the associated TBA system.
We explicitly check 
that  the TBA equations for the $n$-cusp minimal surfaces
are obtained from  the  $\grp{SU}(n-4)_4/\grp{U}(1)^{n-5}$ 
homogeneous sine-Gordon model.
In section 3, we discuss the TBA systems in the single-mass cases in 
relation to the perturbed $\grp{SU}(4)$ diagonal coset and $W$ minimal models.
In section 4, we discuss the expansion of the free energy, and 
derive the leading-order expansion for $n=6$ and $7$.
In section 5, we extend the formalism of the expansion of the T-/Y-functions 
to the AdS$_{4}$ case, and derive the leading-order expansion for $n=6$ and $7$.
Combining those results, we derive the analytic expansion of the remainder
function for $n=6$ and $7$ in section 6.
We also compare the rescaled remainder function for $n=6$ with the
two-loop one. In the appendix, we summarize a computation of a
three-point function in a non-unitary $W$ minimal model. 

\section{TBA equations for minimal surfaces in AdS$_4$}\label{section:TBAmin}
In this section, we review the computation of the  regularized area of  the minimal
surfaces in the AdS space with a null polygonal boundary using integrability.
As studied in  \cite{Alday:2009dv,Alday:2010vh, Hatsuda:2010cc}, such an area is
governed by a set  of non-linear integral equations of the TBA form 
or the associated T-/Y-systems.  Those equations 
coincide with the TBA equations of the homogeneous sine-Gordon model.

\subsection{Functional relations and TBA equations}
The basic idea to compute the area of the minimal surfaces is as follows.
We start with the non-linear sigma model that describes the classical strings in AdS$_5$.
After the Pohlmeyer reduction, the equations of motion for classical
strings are mapped to a linear system of differential equations.
Due to the integrability of the linear system, 
one can introduce a spectral parameter $\theta$.
Using the bispinor representation, this system 
is brought to the $\grp{SU}(4)$ Hitchin system with a $\bbZ_4$-symmetry.
Solutions of this linear problem show the Stokes phenomena \cite{Alday:2009yn}.
The smallest solution is uniquely determined in each Stokes
sector. Their Wronskians evaluated at special values of the spectral
parameter form the cross-ratios of the cusp coordinates.
From  the Pl\"ucker relations, these Wronskians satisfy  
certain functional relations called the T-system. 

For AdS$_5$, 
it reads as the following relations among the T-functions $T_{a,s}(\theta)$,
\begin{align}\label{Tsystem}
T_{a,s}^+ T_{4-a,s}^-=T_{4-a,s+1}T_{a,s-1}+T_{a+1,s}T_{a-1,s}\, ,
\end{align}
where $a=1,2,3$; $s=1,2,\cdots, n-5$  for the $n$-cusp minimal surfaces and 
$f^{\pm}(\theta):=f(\theta\pm {i\pi\over 4})$.
The boundary conditions for the T-functions are
\begin{equation}\label{Tbd1}
T_{a,0}=1\ (a=1,2,3), \quad  T_{0,s}=T_{4,s}=1 \ (s\in \bbZ).
\end{equation}
At the boundary $s=n-4$, we have also to impose the boundary condition related 
to the formal monodromy \cite{Alday:2010vh}.
For $n \not \in 4\mathbb{Z}$, the condition is simply given by 
\begin{align}\label{Tbd2}
T_{1,n-4}= \mu^{-(1+(-1)^{n})/2},
 \quad T_{2,n-4}=1,\quad T_{3,n-4}=\mu^{(1+(-1)^n)/2},
\end{align}
where $\mu$ is a constant.
In this work, we focus on this $n \notin 4 \bbZ $ case, where we do not need
to consider extra monodromy factors.
From the T-functions, the Y-functions are defined by
\begin{align}\label{YTrel}
Y_{a,s}= \frac{T_{a,s+1}T_{4-a,s-1}}{T_{a+1,s}T_{a-1,s}}.
\end{align}
They satisfy a set of functional relations 
called the Y-system:
\begin{align}\label{Ysystem}
\frac{Y_{a,s}^-Y_{4-a,s}^+}{Y_{a+1,s}Y_{a-1,s}}=\frac{(1+Y_{a,s+1})(1+Y_{4-a,s-1})}{(1+Y_{a+1,s})(1+Y_{a-1,s})}.
\end{align}
The boundary conditions for the Y-functions are 
$Y_{a,0} =Y_{a, n-4}=0$ ($a=1,2,3$) and $Y_{0,s}=Y_{4,s}=\infty$ 
 ($s=1,\cdots, n-5$).
The Y-system has many solutions in general.
To determine a solution of the Y-system, we need to know
the analytic structure of the Y-functions including their asymptotics  for
large $ |\theta |$, which has been studied in \cite{Alday:2010vh}.
The asymptotics is specified by auxiliary complex (mass) parameters $m_s$
and constants $C_s, D_s$.  
For  real $m_{s}$, it is given by
\begin{align}\label{Yasympt}
\log Y_{1,s}(\theta)&\to -m_s \cosh \theta-C_s \pm D_s  \comma \nn \\
\log Y_{2,s}(\theta)&\to -\sqrt{2}m_s \cosh \theta \comma \\
\log Y_{3,s}(\theta)&\to  -m_s \cosh \theta+C_s \mp D_s \comma \nn 
\end{align}
as $\theta \to \pm \infty $.
One can show that the Y-system can
be rewritten into a set of  integral equations of the TBA form.
Those equations for the minimal surfaces in the AdS$_5$ space are given by
\begin{align}
\log Y_{1,s}(\theta)&=-m_s \cosh \theta-C_s-K_1*\alpha_s
 -\frac{1}{2}K_2*\beta_s-\frac{1}{2}K_3* \gamma_s , \notag \\ 
\log Y_{2,s}(\theta)&=-\sqrt{2}m_s \cosh \theta-K_2 *\alpha_s-K_1 * \beta_s , 
\label{eq:TBA-AdS5-2}\\
\log Y_{3,s}(\theta)&=-m_s \cosh \theta+C_s-K_1*\alpha_s-\frac{1}{2}K_2*\beta_s+\frac{1}{2}K_3* \gamma_s,  \notag 
\end{align}
where $\ast$ stands for the convolution, 
$f \ast g := \int_{-\infty}^\infty d\theta \, f(\theta-\theta') g(\theta')$.
The functions $\alpha_s$, $\beta_s$ and $\gamma_s$ are defined by
\begin{align}
\alpha_s &= \log \frac{(1+Y_{1,s})(1+Y_{3,s})}{(1+Y_{2,s-1})(1+Y_{2,s+1})},\qquad
\gamma_s=\log \frac{(1+Y_{1,s-1})(1+Y_{3,s+1})}{(1+Y_{1,s+1})(1+Y_{3,s-1})},
 \notag\\
\beta_s &= \log \frac{(1+Y_{2,s})^2}{(1+Y_{1,s-1})(1+Y_{1,s+1})(1+Y_{3,s-1})(1+Y_{3,s+1})}, 
\end{align}
and the kernels are  by
\begin{align}\label{kernels}
K_1(\theta)=\frac{1}{2\pi \cosh \theta},\qquad K_2(\theta)
=\frac{\sqrt{2}\cosh\theta}{\pi \cosh 2\theta},\qquad
K_3(\theta)=\frac{i}{\pi} \tanh 2\theta.
\end{align}
The constants $D_s$ are obtained from  $\gamma_s$ by
$D_s = \frac{i}{\pi} \int d\theta \, \gamma_s (\theta)$, 
whereas the constant $\mu$ in (\ref{Tbd2}) is related to $C_s$.

In this paper, we  particularly focus on the minimal
surfaces in the AdS$_4$ subspace,  
which correspond to the amplitudes with the four-momenta of the external particles 
lying in a three-dimensional subspace. 
In this case, the above integral equations are simplified to the TBA equations of 
a known integrable system.
To reduce the problem into AdS$_4$, we need a projection 
of the original  system for the AdS$_5$ space.
This projection relates the smallest solutions of the linear problem
to those for the inverse problem via a gauge transformation.
This relation results in the following conditions in the TBA system,
\begin{align}\label{13Id}
T_{1,s}(\theta)=T_{3,s}(\theta), \qquad Y_{1,s}(\theta)=Y_{3,s}(\theta),
\end{align}
where the latter relation leads to $\mu^2 =1$.
In this paper, we particularly consider the case with $\mu=1$ and $C_s =0$,
so that one can analyze the area for small $m_s$ by the underlying integrable 
models and conformal field theories. 
Then, we obtain the simplified integral equations,
\begin{align}
\log Y_{1,s}(\theta)&=-m_s \cosh \theta-K_1*\alpha_s-\frac{1}{2}K_2*\beta_s ,
 \nn \\
\log Y_{2,s}(\theta)&=-\sqrt{2}m_s \cosh \theta-K_2 *\alpha_s-K_1 * \beta_s ,  
\label{eq:TBA-1}
\end{align}
where $\alpha_s$ and $\beta_s$ reduce to
\begin{align}
\alpha_s  =\log \frac{(1+Y_{1,s})^2}{(1+Y_{2,s-1})(1+Y_{2,s+1})} ,\qquad
\beta_s =2\log \frac{(1+Y_{2,s})}{(1+Y_{1,s-1})(1+Y_{1,s+1})}.
\end{align}

So far, we have focused on the real mass $(m_s)$ case.
 One can generalize these results  to the complex-mass case as in \cite{Alday:2010vh}.
If the masses in the TBA equations are complex, $m_s=|m_s|e^{i\varphi_s}$, 
the driving terms of the TBA equations are modified as 
\begin{align}
-m_{a,s} \cosh \theta \to -\frac{1}{2}(\bar{m}_{a,s} e^{\theta}+m_{a,s}e^{-\theta})
=-\frac{|m_{a,s}|}{2}(e^{\theta-i\varphi_s}+e^{-(\theta-i\varphi_s)}) ,
\end{align}
where $m_{1,s} = m_{2,s}/\sqrt{2} = m_s$.
Thus, defining $\tilde{Y}_{a,s}(\theta)=Y_{a,s}(\theta+i\varphi_s)$, 
the TBA equations become
\begin{align}
\log \tilde{Y}_{1,s}&=-|m_s| \cosh \theta-2K_1*\log(1+\tilde{Y}_{1,s})\notag \\
&\hspace{0.5cm}+K_1^{s,s-1}*\log(1+\tilde{Y}_{2,s-1})
+K_1^{s,s+1}*\log(1+\tilde{Y}_{2,s+1})  \notag 
\\
&\hspace{0.5cm}-K_2*\log(1+\tilde{Y}_{2,s})+K_2^{s,s-1}*\log(1+\tilde{Y}_{1,s-1})
+K_2^{s,s+1}*\log(1+\tilde{Y}_{1,s+1}) ,\notag \\
\log \tilde{Y}_{2,s}&=-\sqrt{2}|m_s| \cosh \theta-2K_2*\log(1+\tilde{Y}_{1,s})
 \label{eq:TBA-complex1} \\
&\hspace{0.5cm}+K_2^{s,s-1}*\log(1+\tilde{Y}_{2,s-1})
+K_2^{s,s+1}*\log(1+\tilde{Y}_{2,s+1})  \notag \\ 
&\hspace{0.5cm}-2K_1*\log(1+\tilde{Y}_{2,s})+2K_1^{s,s-1}*\log(1+\tilde{Y}_{1,s-1})
+2K_1^{s,s+1}*\log(1+\tilde{Y}_{1,s+1}) ,\notag
\end{align} 
where
\begin{align}
K_j^{s,s'}(\theta)=K_j(\theta+i\varphi_s-i\varphi_{s'}) \qquad (j=1,2).
\end{align}
Note that this integral equations are valid only when $|\varphi_s-\varphi_{s'}|<\pi/4$ for all $s,s'$.
If at least one of $|\varphi_s-\varphi_{s'}|$ is greater than $\pi/4$, we need 
to modify the TBA equations due to the poles in the kernels. 
The complex masses $m_s$ provide $2(n-5)$ independent real parameters 
for the TBA system,
which match the number of the independent cross-ratios formed by the cusp coordinates
of the AdS$_4$ minimal surfaces. 

\subsection{TBA equations for $\grp{SU}(N)_{4}/\grp{U}(1)^{N-1}$ HSG model}

In \cite{Hatsuda:2010cc}, the integral equations for the $2\tilde{n}$-cusp
null-polygonal minimal surfaces in AdS$_{3}$ were identified with the TBA equations 
of the $\grp{SU}(\tilde{n}-2)_{2}/\grp{U}(1)^{\tilde{n}-3}$ HSG  
model with purely imaginary resonance parameters $\sigma_s=i\varphi_s$, 
where the masses of the particles  are regarded as complex parameters. 
The corresponding relation 
was also inferred in the same paper between the present $n$-cusp  minimal surfaces
in AdS$_{4}$ and the $\grp{SU}(n-4)_{4}/\grp{U}(1)^{n-5}$ HSG model. 
The relation is  indeed confirmed  by comparing the integral equations 
in the previous subsection with the TBA equations of this HSG model, which are read 
off from the general expression in \cite{CastroAlvaredo:1999em}.
Let us see this explicitly below. 

For this purpose, we first recall that the HSG model associated with the 
$\grp{SU}(N)$ coset is defined as an integrable perturbation of 
the $\grp{SU}(N)_k/\grp{U}(1)^{N-1}$ gauged WZNW/generalized parafermion
model by the weight-zero
primary fields in the adjoint representation of $\alg{su}(N)$. 
Here, we denote 
the $\grp{SU}(N)$ affine Lie algebra at level $k$ by $\grp{SU}(N)_k$.
This $\grp{SU}(N)_k/\grp{U}(1)^{N-1}$ coset  CFT has the central charge,
\begin{equation}\label{GPFc}
    c({\grp{SU}(N)_k \over \grp{U}(1)^{N-1}})={(k-1)N(N-1)\over k+N} \comma
\end{equation}
and its primary field $\Phi^{\boldsymbol{\Lambda}}_{\boldsymbol{\lambda}}(z)$ 
with  weight $\blambda$ 
 in the highest-weight representation labeled by  $\boldsymbol{\Lambda}$
 has the conformal dimension,
\begin{equation}
\Delta^{\bLambda}_{\blambda}
  ={\bLambda (\bLambda+2\brho_{\alg{su}(N)})\over
 2(k+N)}-{\blambda^2\over 2k}.
\end{equation}
$\brho_{\alg{su}(N)}$  is half the sum of the positive roots
(the Weyl vector) of the Lie algebra $\alg{su}(N)$. 
The action of the HSG model then takes the form,
\eqb\label{HSGaction}
    S_{\rm HSG} = S_{\rm gWZNW} + \lambda \int d^2x \, \Phi \comma 
\eqe
where  $\Phi $ is a combination of the
weight-zero adjoint operators $\Phi^{\bomega_1+\bomega_{N-1}}_{\bf 0}$,
which is parametrized by the $N-1$ real mass
parameters $M_s$ ($s=1,\cdots, N-1$) and the real resonance parameters $\sigma_s$. 
This perturbing operator $\Phi$ has the dimension,
\eqb
   \Delta =  \bar{\Delta} := \Delta^{\bomega_1+\bomega_{N-1}}_{\bf 0}={N\over N+k} \period
\eqe
On dimensional grounds, the coupling of the integrable relevant/mass perturbation 
is expressed by the dimensionless coupling $\kappa$ and the mass scale  $M$ as
\eqb\label{LambdaKappa}
    \lambda = - \kappa M^{2-\Delta - \bar{\Delta}} \period
\eqe
We note that the above action describes a multi-parameter integrable perturbation, 
which is a notable feature of the HSG model.

The particles in this model are labeled by two quantum numbers $(a,s)$ and 
have masses 
\eqb\label{Mratio}
 M_{a,s} = M_{s} \sin\bigl( \frac{\pi a}{k}\bigr)/\sin\bigl( \frac{\pi }{k}\bigr) \comma
\eqe
where $a=1, ..., k-1$. The S-matrix of the diagonal scattering between  the particles 
$(a,r)$ and $(b,s)$ is then given by \cite{Miramontes:1999hx}
\eqb\label{SmatrixHSG}
   S_{ab}^{rs}(\theta;\sigma_{rs}) = 
   \Bigl[ S^{\rm min}_{ab}(\theta)\Bigr]^{\delta_{r,s}} 
    \Bigl[ (\eta_{r,s})^{-ab} S^{F}_{ab}(\theta+\sigma_{rs})\Bigr]^{-I_{rs}} 
    \comma
\eqe
where $\sigma_{rs}:= \sigma_r - \sigma_s$,  and 
\eqb
   S_{ab}^{\rm min}(\theta) =
   \prod_{j=0}^{{\rm min}(a,b)-1} (a+b-2j)_{\theta} (a+b-2-2j)_{\theta} 
\eqe
with
\eqb
     (x)_{\theta} := 
      \frac{\sinh \frac{1}{2}\bigl(\theta + \frac{i\pi}{k} x \bigr)}{
   \sinh \frac{1}{2}\bigl(\theta - \frac{i\pi}{k} x \bigr)} 
\eqe
is the S-matrix of the $A_{k-1}$ minimal affine Toda field theory (ATFT)
\cite{Koberle:1979sg,Braden:1989bu}.  
In the second factor, $I_{rs} = \delta_{r, s+1} + \delta_{r, s-1}$ is
the incidence matrix of the Lie algebra $\alg{su}(N)$, 
$\eta_{r,s} (= \eta_{s,r}^{-1})$  are arbitrary $k^{\rm th}$ roots of $-1$, and
$S_{ab}^{F}$ is given by
\eqb\label{SF}
   S_{ab}^{F}(\theta) = 
   \prod_{j=0}^{{\rm min}(a,b)-1} (a+b-1-2j)_{\theta} \period
\eqe
The parity-invariance of $S_{ab}^{rs}$ is broken due to $\eta_{r,s}$ and 
$\sigma_{rs} $.

Next, we recall that, for a diagonal scattering theory with the S-matrix $S_{AB}$,  
the TBA equations in the fermionic case take the form (see for example \cite{Mussardo}),
\eqb\label{TBA}
    \log Y_{A}(\theta) = -m_{A} \cosh \theta + \sum_{B}  K_{AB} \ast \log (1+Y_{B})
    \comma
\eqe
where $ K_{AB}(\theta)$ are the kernels defined by 
\eqb
   K_{AB}(\theta) = \frac{1}{2\pi i} \frac{\del}{\del \theta} \log S_{AB}(\theta)
   \period
\eqe
On the right-hand side, $m_A = M_A L$ is the dimensionless combination of the 
mass parameter $M_A$ and the length scale/inverse temperature $L$.
We have also assumed  above that the kernels are symmetric:
$ K_{AB}(\theta) = K_{BA}(\theta)$.
Once the resonance parameters are set to be vanishing, $\sigma_{rs} =0$,
the kernels for the  HSG model are indeed symmetric
and one can apply this formula. 

From the resultant TBA equations, one can show that $Y_{1,s} = Y_{3,s}$
for  $k=4$. After imposing this condition, 
we  find that the TBA equations of the $\grp{SU}(N)_{4}/\grp{U}(1)^{N-1}$ HSG model
with vanishing $\sigma_s$ are just the same as the integral equations 
\eqref{eq:TBA-1} for the $(N+4)$-cusp 
minimal surfaces in AdS$_{4}$
with real mass parameters. 
Here, the correspondences of the parameters are
$m_{1,s} = m_{3,s} = m_{2,s}/\sqrt{2} = m_s$ for the masses $m_A = m_{a,s}$, and
\eqb
    K_{12}^{rs} = K_{23}^{rs} = K_1  \comma \quad
   K_{22}^{rs}  =  K_{11}^{rs} +  K_{13}^{rs} = K_{33}^{rs} +  K_{13}^{rs} =  K_2
   \comma
\eqe
for the kernels $K_{AB} = K_{ab}^{rs}$. 
When the  mass parameters are complex,
$m_{s} = |m_{s}|e^{i\varphi_{s}}$, the phases correspond to the purely imaginary 
resonance parameters $\sigma_{s} = i \varphi_{s} $. One  finds 
that the TBA equations
in this case are given by (\ref{eq:TBA-complex1}). 

\subsection{Remainder function}
In the previous two subsections, we have seen that the null-polygonal minimal surfaces 
with $n$ cusps in AdS$_4$ are characterized by the TBA equations 
for the SU($n-4$)$_4$/U(1)$^{n-5}$ HSG model.
We would like to know the area of such minimal surfaces.
Here we see that the area can be expressed in terms of the T-/Y-functions, 
the free energy and the mass parameters associated with the TBA system.

The area shows divergence, since the surfaces extend to the boundary 
of AdS at infinity and have the cusp points there. 
Introducing the radial-cutoff, the regularized area 
is given by the Stokes data of the linear problem.
For the $n$-cusp minimal surfaces, it takes the form,
\begin{equation}
A_n=A_{\rm div}+A_{\rm BDS-like}+A_{\rm periods}+A_{\rm free},
\end{equation}
where $A_{\rm div}$ is the divergent part, $A_{\rm
periods}$ is the period part which depends on the mass parameters governing 
the asymptotics of the Y-functions. $A_{\rm BDS-like}$ 
is given by distances among the cusp points, which is similar to the BDS
expression but different.
$A_{\rm free}$ is the free energy  associated with the TBA system.

The remainder function is now defined by the difference between the regularized 
area  and the BDS formula,
\begin{align}
A_n=A_{\rm div}+A_{\rm BDS}+R_n \period
\end{align}
The explicit form of the remainder function at  strong coupling is then given by
\begin{align}
R_n=\Delta A_{\rm BDS}+A_{\rm periods}+A_{\rm free} \period
\end{align}
The first term $\Delta A_{\rm BDS}:=A_{\rm BDS\mbox{-}like} -
A_{\rm BDS}$ is expressed in terms of the cross-ratios, and
its general expression for $n \notin 4 \bbZ$ is found in \cite{Alday:2009dv}.
Here, we list the expressions for $n=6$ and $7$, which are used
in the following sections:
\eqb\label{delABDSn=6}
  \Delta A_{\rm BDS} ^{(n=6)}
     \Eqn{=} -\frac{1}{4} \sum_{i=1}^{3} \Bigl[ \frac{1}{2} \log^{2}u_{i,i+3} + \dilog(1-u_{i,i+3})\Bigr]
      = \frac{1}{4} \sum_{i=1}^{3} \dilog\Bigl(1- \frac{1}{u_{i,i+3}}\Bigr) \comma 
\eqe
for $n=6$ and 
\eqb\label{delABDSn=7}
   \Delta A_{\rm BDS}^{(n=7)} 
      \Eqn{=} -\frac{1}{4} \sum_{i=1}^{7} \Bigl[ \log^{2}u_{i,i+3} 
      -\frac{1}{2} \log u_{i,i+3} \log\frac{u_{i+2,i+5}u_{i+1,i+5}}{u_{i+3,i+6}u_{i,i+4}}
      + \dilog(1-u_{i,i+3}) \Bigr]
      \comma \qquad 
\eqe
for $n=7$ (see also \cite{Yang:2010as}). 
The cross-ratios $u_{i,j}$ above are defined by
\eqb
   u_{i,j} := \frac{x^{2}_{i,j+1}x^{2}_{i+1,j}}{x^{2}_{i,j} x^{2}_{i+1,j+1}} \comma
\eqe
and the cusps are labeled modulo $n$. These cross-ratios are 
concisely expressed by the Y-/T-functions 
through 
\eqb \label{eq:U-Y}
   U_{s}^{[r]} := 1+ \frac{1}{Y_{2,s}^{[r]}} = \frac{T_{2,s}^{[r+1]} T_{2,s}^{[r-1]}
   }{T_{2,s+1}^{[r]} T_{2,s-1}^{[r]}} 
\eqe
with $f^{[r]}:=f(\theta=i\pi r/4)$  as follows  \cite{Alday:2010vh},
\eqb\label{Uu}
   U_{2k-2}^{[0]}  = \frac{1}{u_{k-1,-k-1}} \comma \qquad 
   U_{2k-1}^{[-1]}  = \frac{1}{u_{k-1,-k-2}}  \period
\eqe
Other cross-ratios are generated by the $\bbZ_{n}$-symmetry, 
$x_{i}^{\mu} \to x_{i+1}^{\mu}$, which corresponds to the shift of
the argument
\eqb
     Y_{a,s}^{[r]} \to  Y_{a,s}^{[r+2]} \comma \qquad  T_{a,s}^{[r]} \to  T_{a,s}^{[r+2]} \period
\eqe

In addition, other parts $A_{\rm periods}$ and $A_{\rm free}$ are given by
\begin{align}
A_{\rm periods}&=\sum_{s,s'=1}^{n-5} {\cal K}_{ss'} m_s \bar{m}_{s'}, \notag\\
A_{\rm free}&=\sum_{s=1}^{n-5}\int_{-\infty}^\infty \frac{d\theta}{2\pi} |m_s| \cosh \theta
\log \Bigl[ (1+\tilde{Y}_{1,s}(\theta))^2 (1+\tilde{Y}_{2,s}(\theta))^{\sqrt{2}} \Bigr].
\end{align}
The explicit forms of ${\cal K}_{ss'}$ are found in \cite{Alday:2010vh} 
for $n\not\in 4\mathbb{Z}$, and  
are conjectured in \cite{Yang:2010az} for $n\in 4\mathbb{Z}$.
Here we  list the results for $n=6,7$ only: 
\begin{align}
{\cal K}^{(n=6)}=\frac{1}{4},\qquad 
{\cal K}^{(n=7)}=\frac{1}{2\sqrt{2}}\begin{pmatrix} \sqrt{2} & 1 \\ 1 & \sqrt{2} \end{pmatrix}.
\end{align}

Although $\Delta A_{\rm BDS}$ is 
given by the cross-ratios directly,  $A_{\rm periods}$ and $A_{\rm free}$ are
related to the cross-ratios indirectly through the Y-/T-functions and the mass parameters.
Indeed,  the Y-functions are uniquely determined by solving the TBA equations
for given masses and, once $Y_{a,s}$ are obtained in terms of $m_s$, 
the mass parameters and hence the Y-/T-functions
are related to the cross-rations  through \eqref{eq:U-Y}.
As a result, the remainder function  at strong coupling is expressed as a function of the
cross-ratios.

\subsection{UV expansion}

In the following sections, we discuss an analytic expansion
of the remainder function around the high-temperature/UV limit, where
the mass parameters $m_s$ become vanishing and the corresponding
Wilson loops become regular-polygonal.
This is achieved by several steps: First, we note that around this limit
the deformation term in (\ref{HSGaction}) is treated as a small-mass
perturbation for the coset/generalized parafermion CFT \cite{Gepner:1987sm}. 
Then,  the free energy of the TBA system,
which is given by the ground-state energy in the mirror channel, is obtained
analytically by the conformal perturbation theory \cite{Zamolodchikov:1989cf}. 
It is expanded in terms of the correlation functions of the deformation operator $\Phi$. 
Next, we use the relation between the Y-/T-function
and  the $g$-function \cite{Affleck:1991tk}. The $g$-function is
regarded as a boundary contribution to the free energy, and analytically 
expanded by the CPT with boundaries \cite{Dorey:1999cj,Dorey:2005ak}.
In the course of the discussion, we first set the mass parameters to be real
to keep the boundary integrability. Their phases are recovered after the expansion 
is obtained, so that the $\bbZ_n$-symmetry is maintained.

These expansions are first given in terms of
the coupling $\lambda$. To find the expansion in terms of the mass parameters,
we need the precise form of $\Phi $ and the relation between $\lambda \Phi$
and $m_s$.  
Once this mass-coupling relation is found, one can obtain the expansion in terms of the
cross-ratios through \eqref{eq:U-Y} as discussed in the previous subsection.

Since there are multiple deformation operators in our case, it is a rather difficult problem
to find the exact mass-coupling relation due to operator mixing. 
However, when some mass parameters are turned off so as to leave only 
one mass scale (single-mass case), 
the TBA system reduces to simpler ones and the problem becomes tractable.

In the next section, we begin our discussion of the UV expansion by
considering  the perturbation with single mass scale  
for the AdS$_{4}$ minimal surfaces.
We see that the TBA systems in such cases  
reduce to those of the perturbed $\grp{SU}(4)$ diagonal coset models or 
 $W$ minimal models. For the 6- and 7-cusp cases ($n=6, 7$), it turns out  that the 
input from the $W$ minimal models is enough to  completely determine  
the leading-order expansion.

\section{Perturbation with single mass scale and $W$ minimal models}\label{section:singlemass}

Before discussing the perturbation with single mass scale
for the AdS$_4$ minimal surfaces,
let us first recall those for the AdS$_3$ case  
\cite{Hatsuda:2011ke,Hatsuda:2011jn}.
The minimal surfaces embedded in AdS$_3$ with $2\tilde{n}$ cusps
are described by the TBA system of the 
$\grp{SU}(\tilde{n}-2)_2/\grp{U}(1)^{\tilde{n}-3}$ HSG model, which is obtained as
the perturbed  $\grp{SU}(\tilde{n}-2)_2/\grp{U}(1)^{\tilde{n}-3}$ generalized 
parafermion
model by the weight-zero $\alg{su}(\tilde{n}-2)$ adjoint operators with dimension
$\Delta = \bar{\Delta} = (\tiln-2)/\tiln$.
The TBA system is characterized by the $A_{\tilde{n}-3}$ Dynkin diagram,
where the  mass parameters $m_s = M_s L$  are associated to each node.
When only one mass parameter  is non-zero, 
$M_s = \delta_{s,r} M$ ($r= 1, ..., \tiln -3$), 
the TBA equations reduce to those for the unitary
$\grp{SU}(2)_r\times \grp{SU}(2)_{\tilde{n}-2-r}/\grp{SU}(2)_{\tilde{n}-2}$ 
  diagonal coset  model perturbed by the
$\phi_{(1,1,{\rm adj})}$ operator with dimension 
$h_{(1,1,{\rm adj})} = \bar{h}_{(1,1,{\rm adj})} = (\tiln-2)/\tiln$ \cite{Zamolodchikov:1991vg}. 
In particular, when $r=1$, 
they become those of the unitary minimal model 
${\cal M}_{\tilde{n}-1,\tilde{n}}$  perturbed by the $\phi_{(1,3)}$ operator
\cite{Itoyama:1990pv,Zamolodchikov:1991vh}.

Furthermore,  when
 the mass parameters are non-vanishing only at a pair of nodes,   
$ M_s = (\delta_{s,r} + \delta_{s, \tiln-2-r}) M $, with $\tiln$ odd, the TBA 
system admits an orbifolding by the $\bbZ_2$-action, and is 
characterized by the $T_{(\tiln-3)/2} = A_{\tiln-3}/\bbZ_2$ tadpole 
diagram with a mass parameter only at the $r^{\rm th}$ node. 
The TBA equations  then reduce to those for the non-unitary
$ \grp{SU}(2)_r \times \grp{SU}(2)_{\tilde{n}/2-2-r}/\grp{SU}(2)_{\tilde{n}/2-2}$ 
 diagonal coset model perturbed
by $\phi_{(1,1,{\rm adj})}$ with dimension 
$h_{(1,1,{\rm adj})} = \bar{h}_{(1,1,{\rm adj})} = (\tiln-4)/\tiln$ \cite{Ravanini:1992fi}.
The exponents of the UV expansion of observables are given by the dimension 
of the the perturbing operator (see the following sections). 
The relation $1-h_{(1,1,{\rm adj})} = 2(1-\Delta)$ assures a consistency between  
the UV expansions from the generalized parafermion and the diagonal coset model,
respectively. In particular, when $r=1$, this  perturbed diagonal coset model 
becomes equivalent to the non-unitary minimal model 
${\cal M}_{\tilde{n}-2,\tilde{n}}$ perturbed by $\phi_{(1,3)}$. 
For  the 10-point remainder function, the results from these unitary and non-unitary 
minimal models and their continuation to complex masses are enough to completely 
determine the leading-order analytic expansion around the UV limit \cite{Hatsuda:2011ke}.

\begin{figure}[t]
 \begin{center}
  \begin{minipage}{0.3\hsize}
  \begin{center}
   \includegraphics[width=50mm]{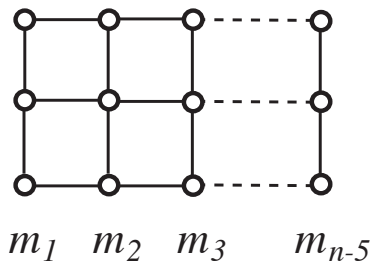}
  \end{center}
 \end{minipage}
  \caption{$A_3 \times A_{n-5}$ diagram of the TBA system for the $n$-cusp  
   minimal surfaces in AdS$_4$.}
    \label{fig:TBADynkin}
  \end{center}
\end{figure}

As discussed in the previous section, 
the $n$-cusp minimal surfaces  in AdS$_4$ 
are described by the TBA system of 
the $\grp{SU}(n-4)_4/\grp{U}(1)^{n-5}$ HSG model, which is obtained as
the perturbed  $\grp{SU}(n-4)_4/\grp{U}(1)^{n-5}$ generalized parafermion
model by the weight-zero $\alg{su}(n-4)$ adjoint operators with dimension
$\Delta = \bar{\Delta} = (n-4)/n$. The TBA system
is characterized by the rectangular diagram  $A_3 \times A_{n-5}$, where one
has the $A_3$ Dynkin diagram in the vertical direction and 
the  $A_{n-5}$ Dynkin diagram in the horizontal direction (Fig.~\ref{fig:TBADynkin}).
The  mass parameters $m_s = M_s L$  are associated to each node of 
the $A_{n-5}$ diagram.  Compared with the AdS$_3$ case, we notice that
the level $k=2$ is replaced with $k=4$ in the AdS$_4$ case.

\subsection{Perturbed unitary diagonal coset/$W$ minimal models}\label{subsection:uniW}

When only one mass parameter is turned on, $M_s = \delta_{s,r} M$, 
one expects from the AdS$_3$ case that 
the TBA system of the HSG model reduces to that for the unitary
\eqb\label{SU4diagcoset}
   \grp{SU}(4)_r\times \grp{SU}(4)_{n-4-r}/\grp{SU}(4)_{n-4}
\eqe
 diagonal coset model perturbed by $\phi_{(1,1,{\rm adj})}$ with dimension 
$h_{(1,1,{\rm adj})} = \bar{h}_{(1,1,{\rm adj})} = (n-4)/n$. In particular,
when $r=1$, the above model becomes equivalent to 
the perturbed unitary $W$ minimal model, 
\eqb
   WA_3^{(n-1,n)} \period
\eqe    
Here, we have used the relations (\ref{WApqCoset}) and (\ref{WApqCosetLevel}).

This expectation is also supported by 
an observation that the TBA system of the $\grp{SU}(N)$ Gross-Neveu
model, which is characterized by the $A_{N-1} \times A_{m}$ diagram 
with $ m \to \infty$, 
is given by the TBA system of the perturbed 
$\grp{SU}(N)_1\times \grp{SU}(N)_{m}/\grp{SU}(N)_{1+m}$
diagonal coset model  \cite{Fendley:2001hc}. 
Indeed, one can explicitly check that the above correspondences of the TBA
systems are correct by comparing the TBA equations of the HSG model and 
those  of the 
$\grp{G}_k \times \grp{G}_k/\grp{G}_{k+l}$ diagonal coset  model perturbed by 
$\phi_{(1,1,{\rm adj})}$ \cite{Ravanini:1992fs}.

\subsection{Perturbed non-unitary  diagonal coset/$W$ minimal models}
\label{subsection:nonuniW}

When a pair of the mass parameters are turned on, 
$M_s = (\delta_{s,r}+\delta_{s,n-4-r}) M$, with $n$ odd, the TBA 
system is characterized by  the diagram $(A_3 \times T_{(n-5)/2})_r $, namely,
the $A_3 \times T_{(n-5)/2} $
diagram with a mass parameter only for the $r^{\rm th}$ column. 
Taking into account the above and AdS$_3$ cases, 
one then expects  that 
the TBA system in this case reduces to that for the non-unitary
\eqb\label{SU4diagcoset2}
   \grp{SU}(4)_r\times \grp{SU}(4)_{n/2-4-r}/\grp{SU}(4)_{n/2-4}
\eqe
 diagonal coset model perturbed by $\phi_{(1,1,{\rm adj})}$
with dimension 
$h_{(1,1,{\rm adj})} = \bar{h}_{(1,1,{\rm adj})} = (n-8)/n$. 
The relation $1-h_{(1,1,{\rm adj})} = 2(1-\Delta)$  is consistent with the UV expansion.
In particular, when $r=1$, this perturbed diagonal coset model 
becomes equivalent to the perturbed non-unitary $W$ minimal model,  
\eqb
   WA_3^{(n-2,n)} \period
\eqe    

These are particular examples of the correspondence 
between the TBA system characterized by  the diagram $( \grp{G} \times T_{l})_r $  
and the non-unitary diagonal coset
model for $\grp{G}$, which has been suggested in \cite{Ravanini:1992qb}. 
In the following sections, assuming, in particular, the correspondence 
for the 7-cusp ($n=7$) case,
we derive the analytic expansion of the remainder function around the UV limit, to
find a good agreement with the results from the numerical computation. We regard
this also as a non-trivial check of the above correspondence.

\subsection{$W$ minimal models}
In the previous subsections, we observed/argued that the TBA systems for the
AdS$_4$ minimal surfaces in the single-mass cases
 reduce to those for the diagonal coset/$W$ minimal models. 
As mentioned,
we consider the cases corresponding to the $W$ minimal models 
to determine the  UV expansion of the remainder function for $n=6$ and $7$.
For later use, we thus summarize the $W$ minimal model below. 

In the following, we focus on  the $WA_{k-1}^{(p,q)}$ minimal model \cite{Fateev:1987zh}, 
where $p,q$ ($p < q$) are positive and relatively prime integers.
The central charge of the model is
\eqb\label{cWA}
  c(WA_{k-1}^{(p,q)})=(k-1)\left(1-{k(k+1) (p-q)^2\over pq}\right) \period
\eqe
The primary fields $\Phi_{l,l'}$ have the dimensions
\eqb\label{hll}
  h_{l,l'}={12\boldsymbol{\Lambda}_{l,l'}^2-k(k^2-1)(p-q)^2\over 24 pq} \period
\eqe
Here,  $l=(l_1,\cdots, l_{k-1})$ and $l'=(l'_1,\cdots, l'_{k-1})$ 
are vectors of positive integers satisfying 
\eqb\label{llqp}
\sum_{i=1}^{k-1}l_i\leq q-1,\quad 
\sum_{i=1}^{k-1}l'_i\leq p-1 \period
\eqe
$\boldsymbol{\Lambda}_{l,l'}$ is given by
\eqb
\boldsymbol{\Lambda}_{l,l'}=\sum_{i=1}^{k-1}(p l_i-q l'_i) \boldsymbol{\omega}_i \comma
\eqe
where $\boldsymbol{\omega}_{i}$ ($i=1,\cdots,k-1$) are 
the fundamental weights of  $A_{k-1}$
normalized as 
\eqb
\boldsymbol{\omega}_i\cdot \boldsymbol{\omega}_j
={i(k-j)\over k} \quad {\rm for} \quad  i\leq j \period
\eqe
We also define the effective central charge by
$c_{\rm eff}(WA_{k-1}^{(p,q)}) :=c(WA^{(p,q)}_{k-1})-24 h_0$, 
where $h_0$ denotes the lowest conformal weight.
For  the unitary model with $q=p+1$, the lowest weight is 0,
but otherwise it is evaluated as \cite{KacWakimoto}
\begin{equation}
c_{\rm eff}(WA_{k-1}^{(p,q)})=(k-1)\left(1-{k(k+1)\over pq}\right) \period
\label{eq:effc}
\end{equation}

The $WA_{k-1}^{(p,q)}$ minimal model is represented by the coset model as  
\cite{Frenkel:1992ju}
\eqb\label{WApqCoset}
    WA_{k-1}^{(p,q)} = {\grp{SU}(k)_{1} \times \grp{SU}(k)_{m} \over \grp{SU}(k)_{1+m} }
    \comma
\eqe
 where
\eqb\label{WApqCosetLevel}
  m +k = \frac{p}{q-p} \period
\eqe
We note  that $m$ 
 is not generally a non-negative integer corresponding to
an integrable representation. Instead, the general $m$ corresponds to
an admissible representation. 
Let $(\boldsymbol{\mu}_{1},\boldsymbol{\mu}_{m},\boldsymbol{\mu}_{m+1})$ be 
the weighs of $\alg{su}(k)$ for 
$\grp{SU}(k)_{1}, \grp{SU}(k)_{m}$ and $\grp{SU}(k)_{m+1}$,
respectively. Since $\boldsymbol{\mu}_{1}$ 
is determined by other two weights 
\cite{KacWakimoto,Frenkel:1992ju,Mathieu:1991fz}, one can label
the fields in the coset model by
\eqb
   (\boldsymbol{\Lambda}_{+}, \boldsymbol{\Lambda}_{-}) 
:= (\boldsymbol{\mu}_{m},\boldsymbol{\mu}_{m+1}) \period
\eqe
Then, the dimension of the field is given by
\eqb\label{h+-}
   h_{(\boldsymbol{\Lambda}_{+},\boldsymbol{\Lambda}_{-})} 
= { \bigl[ q \boldsymbol{\Lambda}_{+} -p\boldsymbol{\Lambda}_{-} 
   + (q-p) \boldsymbol{\rho} \bigr]^{2} 
   -(q-p)^{2}\boldsymbol{\rho}^{2} \over 2 pq}
   \comma
\eqe
where $\boldsymbol{\rho}$ is the Weyl vector of 
$\alg{su}(k)$, i.e., 
\eqb
   \boldsymbol{\rho} = \sum_{i=1}^{k-1} \boldsymbol{\omega}_{i} \period
\eqe
Since $\boldsymbol{\rho}^{2} = k(k^{2}-1)/12$, comparing (\ref{h+-}) and (\ref{hll}) gives
\eqb\label{ll+-}
   \pm \boldsymbol{\Lambda}_{l,l'} = q\boldsymbol{\Lambda}_{+} 
 -p \boldsymbol{\Lambda}_{-} +(q-p) \boldsymbol{\rho} \comma
\eqe
up to field identifications.

For example, the perturbing operator $\phi_{(1,1,{\rm adj})}$ for  the single-mass cases
is labeled by 
\eqb\label{w11ad}
 (\boldsymbol{\Lambda}_{+},\boldsymbol{\Lambda}_{-}) 
= (0, \boldsymbol{\omega}_{1}+\boldsymbol{\omega}_{k-1})
 \comma 
\eqe 
and has the dimension 
\eqb\label{h11ad}
   h_{\rm (1,1,adj)} = \frac{p-(k-1)(q-p)}{q} = \frac{m+1}{m+k+1}\period
\eqe
In addition, for the non-unitary model $ WA_{3}^{(n-2,n)}$
with $n$ odd,  which is used later, 
the vacuum or ground-state operator $\phi_0$  is labeled by
\eqb\label{gs}
     (\boldsymbol{\Lambda}_{+},\boldsymbol{\Lambda}_{-}) =
     \Bigl (\frac{n-7}{2}\boldsymbol{\omega}_{2}, \frac{n-5}{2}\boldsymbol{\omega}_{2} \Bigr)
 \comma
\eqe
and has the dimension
 \begin{equation}
h_0=-{15\over 2n(n-2)} \period
\label{h0}
\end{equation} 
The effective central charge  is then
$
 c_{\rm eff}=3 \bigl(1-{20\over n(n-2)} \bigr)\period 
$
\subsection{Level-rank duality  and decomposition of coset models}

In subsection \ref{subsection:uniW}, we discussed the relation 
between the TBA systems of the 
HSG model in the single-mass cases and those of the perturbed unitary $W$
minimal models.  This relation is directly found by using 
 a decomposition of the generalized parafermion model
into a product of the diagonal coset models based on 
the level-rank duality \cite{Altschuler:1988mg,Kuniba:1990zh}.

Let us start with a simple example of the  $\grp{SU}(2)_k/\grp{U}(1)$ coset or
 the $\bbZ_k$-parafermion theory \cite{Fateev:1985mm,Gepner:1986hr}, 
 which has the central charge $c=2(k-1)/(k+2)$ according to (\ref{GPFc}). 
 The perturbing operator,   i.e., weight-zero adjoint operator, 
has the conformal dimension $2/(k+2)$.
 By the level-rank duality, this parafermion CFT is equivalent to 
 the $\grp{SU}(k)_1\times \grp{SU}(k)_1 / \grp{SU}(k)_2$ diagonal coset CFT
or the $WA_{k-1}^{(k+1,k+2)}$
minimal model which has the same central charge \cite{Bais:1987zk}.
The perturbing field on the dual side is $\phi_{(1,1,{\rm adj})}$ 
with the dimension  (\ref{h11ad}) for $m=1$, which indeed coincides with $2/(k+2)$. 

The above equivalence between the perturbed 
parafermion and diagonal coset/$W$ minimal models 
can be generalized to the case of  $\grp{SU}(N)_k/\grp{U}(1)^{N-1}$.
We see that 
\eqb
    {\grp{SU}(N)_k\over \grp{U}(1)^{N-1}} 
    \Eqn{=}  {(\grp{SU}(k)_1)^{N}\over \grp{SU}(k)_N} \nn \\
     \Eqn{=} {\grp{SU}(k)_1\times \grp{SU}(k)_1\over \grp{SU}(k)_2} \times
{\grp{SU}(k)_1\times \grp{SU}(k)_2\over \grp{SU}(k)_3}
\times \cdots \times
{\grp{SU}(k)_1\times \grp{SU}(k)_{N-1}\over \grp{SU}(k)_N} \period \quad 
\label{eq:decomp1}
\eqe
The first equation is due to \cite{Bagger:1988px}, and the second expression
is due to \cite{Bouwknegt:1992wg}.
The matching of the central charges follows from
\begin{equation}
 c({\grp{SU}(N)_k \over \grp{U}(1)^{N-1}}) - c({\grp{SU}(N-1)_k \over \grp{U}(1)^{N-2}})
  =c(WA_{k-1}^{(k+N-1,k+N)}). 
\end{equation}
In the decomposition (\ref{eq:decomp1}), the dimension of  
the perturbing operator on the left-hand side 
is $\Delta = N/(N+k)$, which coincides  with  that of $\phi_{(1,1,{\rm adj})}$ in the rightmost 
model on the right-hand side. This means that a weight-zero operator 
on the l.h.s. is represented solely by the $\phi_{(1,1,{\rm adj})}$ operator in 
the rightmost model, and thus the corresponding single-mass case is described by 
the  $ {\grp{SU}(k)_1\times \grp{SU}(k)_{N-1}/ \grp{SU}(k)_N} 
= WA_{k-1}^{(N+k-1,N+k)}$  model perturbed by $\phi_{(1,1,{\rm adj})}$.

For example,  for the level $k=2$ corresponding to the AdS$_{3}$ case, 
one has a product of the unitary minimal models in  (\ref{eq:decomp1}), 
and the relation has also been 
confirmed by the decomposition of the characters \cite{Ninomiya:1986dp,Kac:1988}.
Further setting $N=\tiln -2$ for the $2\tiln$-cusp minimal surfaces, 
the rightmost model becomes the unitary ${\cal M}_{\tiln-1,\tiln}$
minimal model, as already discussed. For $k=4$ and $N=n-4$,
we indeed have the unitary $WA_3^{(n-1,n)}$ minimal model.

One finds a similar ``decomposition" also 
for the TBA system characterized by the $A_{k-1} \times T_{(N-1)/2}$ diagram
with $N$ odd. To see this, we first note the  relation among the central charges,
\begin{equation}\label{relCs}
  c({\grp{SU}(N)_k \over \grp{U}(1)^{N-1}})- c({\grp{SU}(N-2)_k \over \grp{U}(1)^{N-3}})
 =2 c_{\rm eff}(WA_{k-1}^{(k+N-2,k+N)}) \comma
\end{equation}
and then denote the relation after a successive use of it by
\eqb
 {\grp{SU}(N)_k\over \grp{U}(1)^{N-1}}\sim (WA_{k-1}^{(k+1,k+3)})^2
 \star (WA_{k-1}^{(k+3,k+5)})^2  \star \cdots \star    (WA_{k-1}^{(k+N-2,k+N)})^2  .
\label{eq:dec1}
\eqe
In parallel with the decomposition (\ref{eq:decomp1}), we find that 
the rightmost factor on the r.h.s., $WA_{k-1}^{(k+N-2,k+N)} 
= {\grp{SU}(k)_1\times \grp{SU}(k)_{(k+N)/2-k-1}/ \grp{SU}(k)_{(k+N)/2-k}}$, 
 is the non-unitary diagonal coset/$W$ minimal model describing 
 the TBA system in the single-mass case which is characterized 
 by $(A_{k-1} \times T_{(N-1)/2})_1$.
 In particular, for the $n$-cusp minimal surfaces in AdS$_4$ with $n$ odd, we have 
 $WA_{3}^{(n-2,n)} = {\grp{SU}(4)_1\times \grp{SU}(4)_{n/2-5}/ \grp{SU}(4)_{n/2-4}}$, 
 as already observed in subsection \ref{subsection:nonuniW}.

We note that, in the rank 2 cases with $N=3$, there is  only one factor of 
$(WA_{k-1}^{(k+1,k+3)})^2$ on the r.h.s. of (\ref{eq:dec1}), 
which means that the central charge
of the model on the l.h.s. is twice that on the r.h.s.. This is in accord with 
the fact  that the free energy for the $A_{k-1} \times A_2$
TBA system with equal mass parameters is twice that for $(A_{k-1} \times T_{1})_1$
(see the next section). Explicitly,  in the AdS$_{3}$ case with $\tiln=N+2=5$, 
the relation (\ref{eq:dec1}) reads as 
${\grp{SU}(3)_2/ \grp{U}(1)^2}\sim  (WA_1^{(3,5)})^2 = ({\cal M}_{3,5})^2$.
This was used to determine the UV expansion of the remainder function 
for the  10-cusp minimal surfaces \cite{Hatsuda:2011ke}. In the next section, we use the 
relation for the AdS$_4$ case with $n=N+4=7$,
\eqb
    {\grp{SU}(3)_4 \over \grp{U}(1)^3}\sim  (WA_3^{(5,7)})^2  \comma
\eqe
to determine the UV expansion  
for the  7-cusp minimal surfaces. 

The ``decomposition"  (\ref{eq:dec1}) based on the counting of the central charges 
tells us which $W$ minimal model appears
in the single-mass case. It would be of interest to substantiate this relation 
at a more fundamental level.

\section{UV expansion of free energy}\label{section:UVfree}
As explained in section \ref{section:TBAmin}, 
the HSG model is regarded as an integrable perturbation 
of the generalized parafermion theory.
Near the UV fixed point, we can thus analyze it by using the 2d CFT technique.
In this section, we consider the UV expansion of the free energy for the 
$\grp{SU}(n-4)_4/\grp{U}(1)^{n-5}$ HSG model.
In particular, we write down the analytic expression of the UV expansion for $n=6,7$.
The connection between the generalized parafermions and the $W$ minimal models 
in the previous section is useful.
The expansion of the T-functions will be considered in the next section.
Before proceeding  to detailed analysis, 
we note  our notation for the mass parameters: 
\begin{align}
m_s= M_s L = \tilde{M}_s l, \qquad l=ML,
\end{align}
where $\tilde{M}_s$ are the relative masses, $M$ is the overall mass scale, and $L$ is the circumference
of the cylinder on which the HSG model is defined.

Since the weight-zero adjoint operators of the $\grp{SU}(n-4)_4/\grp{U}(1)^{n-5}$ generalized parafermion theory have the dimension $\Delta=\bar{\Delta}=(n-4)/n$,
the free energy is expanded around the UV fixed point $l=0$ as
\cite{Zamolodchikov:1989cf}
\begin{align}
A_{\rm free}=\frac{\pi}{6}c_n +f_n^{\rm bulk}+\sum_{p=2}^\infty f_n^{(p)}l^{8p/n},
\end{align}
where $c_n$ is the central charge and $f_n^{\rm bulk}$ is  the bulk contribution. 
In the case of our interest, $\grp{SU}(n-4)_4/\grp{U}(1)^{n-5}$, the central charge is given by
\begin{align}
c_n=\frac{3(n-4)(n-5)}{n}.
\end{align}
The general form of the bulk term is not known. 
Here we assume, as in the AdS$_3$ case \cite{Hatsuda:2011ke,Hatsuda:2011jn}, 
that this term just cancels the period term $A_{\rm periods}$ around the UV limit,  i.e., 
\begin{align}
f_n^{\rm bulk}=-A_{\rm periods}.
\end{align}
This is equivalent to requiring that the remainder function is expanded 
by $l^{4p/n}$ for $n \notin 4 \bbZ$, as is the case for the T-/Y-functions discussed 
in the next section. For $n=6$, this is indeed the case \cite{Hatsuda:2010vr}, and 
we argue below that this holds also for $n=7$. 
We expect it to be true for any $n \not\in 4\mathbb{Z}$.

The expansion coefficients  $f_n^{(p)}$ are obtained 
from the connected $n$-point correlation functions of the perturbing operator $\Phi$
at the CFT point. In particular, $f_n^{(2)}$ is given by
\begin{align}
f_n^{(2)}=\frac{\pi}{6}(\kappa_n G(\tilde{M}_s))^2 C_n^{(2)},
\end{align}
where we have denoted the dimensionless coupling in (\ref{LambdaKappa})
by $\kappa_n$.
The function $G(\tilde{M}_s)$ is introduced as the normalization of 
the two-point function of 
the perturbing operator $\Phi$ in (\ref{HSGaction}) parametrized by  $\tilde{M}_s$, 
\begin{align}
   \langle \Phi (z) \Phi(0) \rangle
=\frac{G^2(\tilde{M}_s)}{|z|^{4\Delta}} ,
\end{align}
and $C_n^{(2)}$ is given by
\begin{align}
C_n^{(2)}=3(2\pi)^{2-16/n} \gamma^2\( 1-\frac{4}{n}\) \gamma\(\frac{8}{n}-1\) ,
\label{eq:C_n^(2)}
\end{align}
with $ \gamma(x)= \Gamma(x)/\Gamma(1-x)$.
We still need to determine the function $G(\tilde{M}_s)$. 
As discussed below,  it is trivial for $n=6$, whereas for $n=7$ it is determined 
by using the relation between the TBA system and the $W$ minimal models 
in the previous section.

\subsection{Case of six-cusp minimal surfaces ($n=6$)}
In this case, there is only one mass scale. 
Thus the above function is trivially given by $G(\tilde{M}_1) = \tilde{M}_1$, which 
is equal to 1 for real $m_1$. 
As discussed in the previous section, the HSG model for $n=6$ is equivalent 
to a perturbed $\bbZ_4$-parafermion or 
$\grp{SU}(4)_{1} \times \grp{SU}(4)_{1}/\grp{SU}(4)_{2} 
=WA_{3}^{(5,6)} $ model.
The constant $\kappa_6$ is thus read from the exact mass-coupling relation in
\cite{Fateev:1993av}
\begin{align}\label{kappa6}
\kappa_6 G =\frac{1}{2\pi} \gamma^{1/2}\( \frac{1}{6}\) \left[ \sqrt{\pi} \gamma\(\frac{3}{4}\) \right]^{4/3}.
\end{align}
We thus obtain
\begin{align}\label{f62}
f_6^{(2)}=\frac{\pi}{6}\kappa_6^2 G^2 C_6^{(2)}=\frac{\pi}{2}\gamma^3\(\frac{1}{3}\)\gamma\(\frac{1}{6}\)
\left[ \frac{1}{2\sqrt{\pi}} \gamma\(\frac{3}{4}\) \right]^{\frac{8}{3}}.
\end{align} 
This is indeed obtained by setting $\mu = 1$ in the results in \cite{Hatsuda:2011ke}. 
As discussed in \cite{Hatsuda:2011ke}, 
this  expression is continued to the complex-mass case 
as $G^2(\tilde{M}_s) \to |G(\tilde{M}_s e^{i\varphi_s})|^2$ so as to maintain the
$\bbZ_n$-symmetry. The continuation in this case is, however,  trivial, to give $|G|^2 =1$.

\subsection{Case of seven-cusp minimal surfaces ($n=7$)}\label{subsection:Fn=7}
In this case, there are two mass parameters $(m_1,m_2)$,  which we first set to be real.
To fix the function $G(\tilde{M}_1,\tilde{M}_2)$, we use the strategy explained 
in the previous section (see also \cite{Hatsuda:2011ke}).
From the symmetry and the dimensional analysis, 
we see that this function takes the form
\begin{align}\label{GM}
G(\tilde{M}_1,\tilde{M}_2)=\sum_{r,s=1}^2 F_{rs} \tilde{M}_r^{4/7} \tilde{M}_{s}^{4/7},
\end{align} 
where $F_{11}=F_{22}$ and $F_{12}=F_{21}$.
We would like to fix such coefficients.
For this purpose, we consider the following two cases.

Let us first consider the case where $(m_1,m_2) \to (l,0)$. 
In this case, the TBA equations reduce to those for an integrable perturbation of the 
$W$ minimal model, 
\begin{align}
WA_3^{(6,7)}=\grp{SU}(4)_1 \times \grp{SU}(4)_2/\grp{SU}(4)_{3}.
\end{align}
The perturbing operator is the relevant operator $\Phi$ 
with dimension $\Delta=\bar{\Delta}=3/7$.
The bulk term in this TBA system is \cite{Dunning:2002cu}
\eqb\label{bulk1}
   f^{\rm bulk} = - \frac{1}{2} l^{2} \period
\eqe
We can also read off the mass-coupling relation for this perturbed model 
from \cite{Fateev:1993av}:
\begin{align}
\kappa_7 G(1,0)=\kappa_7 F_{11}=\frac{2}{3\pi} \left[ \gamma\( \frac{2}{7}\) 
\gamma\( \frac{4}{7} \) \right]^{1/2}
\left[ \frac{3}{4\sqrt{2}} \Gamma^2\( \frac{3}{4} \) \right]^{8/7}.
\label{eq:kappaF11}
\end{align}

To fix $F_{12}$, let us next consider the case with $m_1=m_2$. 
 As argued in the previous section, 
the TBA equations in this case may be equivalent to those
for an integrable perturbation of the non-unitary $W$ minimal model,
\begin{align}
WA_3^{(5,7)}=\grp{SU}(4)_1 \times \grp{SU}(4)_{-3/2}/{\rm SU(4)}_{-1/2}.
\end{align}
The central charge and the effective central charge of this CFT are given, respectively, by
\begin{align}
c(WA_3^{(5,7)})=-\frac{27}{7},\qquad
c_{\rm eff}(WA_3^{(5,7)})=\frac{9}{7} .
\end{align}
The perturbing operator $\hat{\Phi} = \phi_{(1,1,{\rm adj})}$ 
labeled by the weight (\ref{w11ad}) has the conformal dimension  
$\hat{\Delta} = \bar{\hat{\Delta}}=-1/7$, while 
the vacuum operator $\hat{\Phi}_0 = \phi_0$   labeled by (\ref{gs})  
has the  dimension
$ \Delta_0=\bar{\Delta}_0=-{3}/{14} $.

Now, let us consider the UV expansion of the free energy for this TBA system.
The mass-coupling relation \cite{Fateev:1993av} reads as
\begin{align}
\hat{\lambda}=\hat{\kappa}M^{32/7},
\end{align}
where
\begin{align}
(\pi \hat{\kappa})^2=9 \gamma\(\frac{1}{7}\)
\gamma\(-\frac{3}{7}\) \left[
\frac{\Gamma(\frac{3}{4})\Gamma(\frac{7}{8})}{2\Gamma(\frac{5}{8})} 
\right]^{\frac{32}{7}} .
\end{align}
The free energy is expanded around the UV fixed point as
\begin{align}
\hat{F}(l)=\frac{\pi}{6}c_{\rm eff}+   \hat{f}^{\rm bulk}(l) 
+\sum_{p=1}^\infty \hat{f}^{(p)} l^{16p/n} ,
\end{align}
where the bulk term is given by \cite{Dunning:2002cu}
\begin{align}\label{bulk2}
 \hat{f}^{\rm bulk}(l) 
=-\frac{1+\sqrt{2}}{2\sqrt{2}}l^2,
\end{align}
and the coefficients $\hat{f}^{(p)}$ are expressed as
\begin{align}
\hat{f}^{(p)}=\frac{\pi}{6}\hat{\kappa}^p \hat{C}^{(p)}.
\end{align}
The coefficients $\hat{C}^{(p)}$ are given by the correlation functions
of the vacuum and the perturbing operators, and the
integral forms of $\hat{C}^{(p)}$ are found in \cite{Zamolodchikov:1989cf,Klassen:1990dx}.
Here, we are interested in the first correction given by 
\begin{align}
\hat{f}^{(1)}=\frac{\pi}{6}\hat{\kappa}\hat{C}^{(1)},\qquad
\hat{C}^{(1)}=-12(2\pi)^{2 \hat{\Delta}-1}C_{\hat{\Phi}_0\hat{\Phi}\hat{\Phi}_0} ,
\end{align}
where $C_{\hat{\Phi}_0\hat{\Phi}\hat{\Phi}_0}$ is the three-point structure constant.
This structure constant is computed in Appendix A, and given by  \eqref{eq:SC}.
Thus the first correction is finally given by
\begin{align}
\hat{f}^{(1)}=\frac{1}{2\pi^{9/7}}\gamma\(\frac{2}{7}\)\gamma\(\frac{1}{14}\)
\left[\frac{\Gamma(\frac{3}{4})\Gamma(\frac{7}{8})}{2\Gamma(\frac{5}{8})} \right]^{\frac{16}{7}}.
\label{eq:hatf1}
\end{align}

From this result, we can fix $F_{12}$.
For $m_1=m_2$ ($\tilde{M}_1=\tilde{M}_2=1$),  we find from (\ref{GM}) that
\begin{align}
f_7^{(2)}
=\frac{\pi}{6} (\kappa_7 F_{11})^2C_7^{(2)} \times 4\(1+\frac{F_{12}}{F_{11}}\)^2 .
\end{align}
This correction must be twice   $\hat{f}^{(1)}$.
Using \eqref{eq:C_n^(2)}, \eqref{eq:kappaF11} and \eqref{eq:hatf1}, we thus obtain
\begin{align}
1+\frac{F_{12}}{F_{11}}=\(\frac{\pi^8}{2^3 \cdot 3^2}\)^{\frac{1}{14}}\left[\gamma\(\frac{4}{7}\)\gamma\(\frac{6}{7}\)
\gamma\( \frac{1}{14}\)\right]^{1/2} \left[ \frac{\Gamma(\frac{7}{8})}{\Gamma(\frac{3}{4})
\Gamma(\frac{5}{8})}\right]^{\frac{8}{7}}.
\end{align}

In summary, for $n=7$, the function $G(\tilde{M}_1,\tilde{M}_2)$ has the following form,
\begin{align}\label{kappa7G}
\kappa_7 G(\tilde{M}_1,\tilde{M}_2)=\kappa_7F_{11} (\tilde{M}_1^{8/7}+\tilde{M}_2^{8/7}+B \tilde{M}_1^{4/7} \tilde{M}_2^{4/7}) ,
\end{align}
where $\kappa_7 F_{11}$ is given by \eqref{eq:kappaF11} and the constant $B$ is given by
\begin{align}
B=\(\frac{2^{11}\pi^8}{3^2}\)^{\frac{1}{14}}\left[\gamma\(\frac{4}{7}\)\gamma\(\frac{6}{7}\)
\gamma\( \frac{1}{14}\)\right]^{1/2} \left[ \frac{\Gamma(\frac{7}{8})}{\Gamma(\frac{3}{4})
\Gamma(\frac{5}{8})}\right]^{\frac{8}{7}}-2.
\end{align}
We also find that the bulk terms in the above two cases, (\ref{bulk1}) and (\ref{bulk2}),
fix the form in the general case to be $ f^{\rm bulk} (l) = - A_{\rm periods}$, as expected.
Thus, from the  connections to the $W$ minimal models, we indeed find
this relation for $n=7$.

So far,  we have considered the case of real $m_{s}$.
Let us now consider the UV expansion of the free energy when the masses are complex.
By the relation to $A_{\rm periods}$, 
or  the argument in \cite{Hatsuda:2011ke} to maintain the $\bbZ_{n}$-symmetry,
the bulk term is given by
\begin{align}
  f^{\rm bulk} (l) 
  &=  -A_{\rm periods}=
-\frac{1}{2}(m_1\bar{m}_1+m_2\bar{m}_2)-\frac{1}{2\sqrt{2}}(m_1 \bar{m}_2+m_2 \bar{m}_1) \nonumber \\
&=-\frac{1}{2}\(\tilde{M}_1^2+\tilde{M}_2^2+\sqrt{2}\tilde{M}_1\tilde{M}_2 \cos(\varphi_1-\varphi_2)
\)l^2 ,
\end{align}
where $|m_s|=\tilde{M}_s l$. Similarly, following  \cite{Hatsuda:2011ke}, we find that 
the function $G$ is continued as
\begin{align}
G^2(\tilde{M}_1,\tilde{M}_2) \to  |G(\tilde{M}_1 e^{i\varphi_1},\tilde{M}_2 e^{i\varphi_2})|^2 .
\end{align}

\begin{figure}[t]
 \begin{minipage}{0.49\hsize}
  \begin{center}
   \includegraphics[width=80mm]{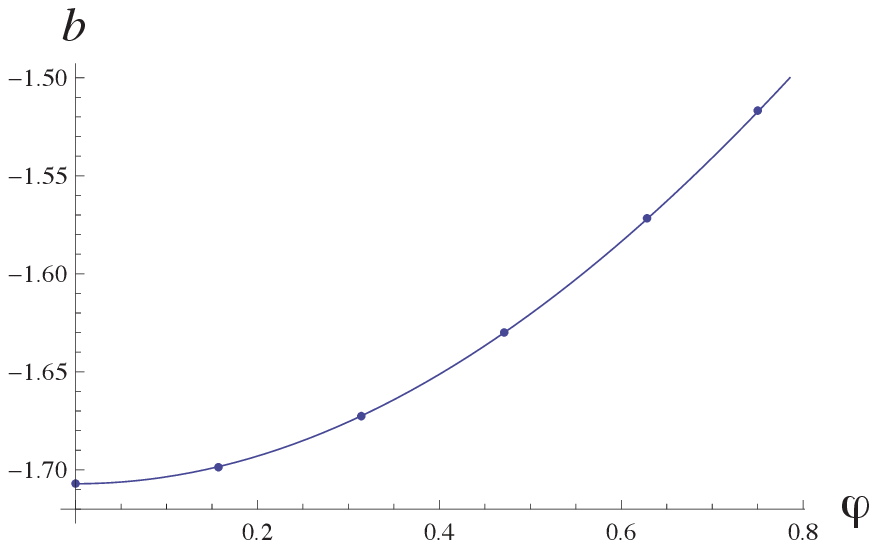} 
  \end{center}
 \end{minipage}
 \begin{minipage}{0.49\hsize}
  \vspace{-2mm}
  \begin{center}
   \includegraphics[width=80mm]{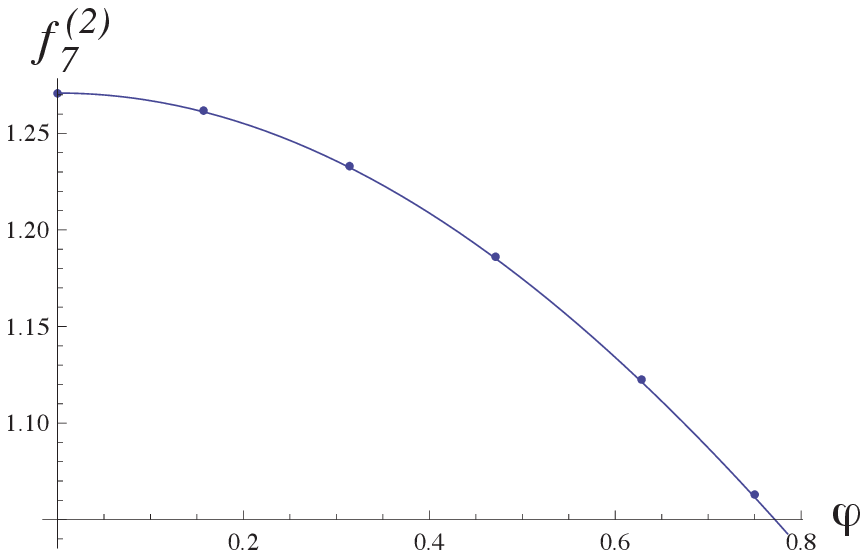} 
  \end{center}
  \end{minipage}
 \caption{The (relative) phase dependence of the bulk term (left) and $f_7^{(2)}$ (right).}
  \label{fig:phi-dep}
\end{figure}

In order to check the validity of these expressions, we compare them with  
numerical results from the TBA equations.
Here, we remark that the TBA equations for the AdS$_{4}$ minimal surfaces
generally exhibit an instability \cite{CastroAlvaredo:2004qi} around the UV limit, 
in that simple iterations for numerics do not converge.
However, we have found that numerics based on the iteration works
if $|m_1|=|m_2|$ up to some value of $l$.  When the phases are turned off,
the numerics works for smaller $l$. For the comparison, 
we have thus solved  the TBA equations for $|m_1|=|m_2|$ from $l=1/20$ to $l=1/2$ 
with step $1/100$ for various values of $\varphi :=\varphi_1-\varphi_2$.
We  have then  fitted the free energy by the function,
\begin{align}
A_{\rm free}^{(\rm fit)}=\frac{\pi}{6}c+b l^2+f_7^{(2)} l^{16/7}+f_7^{(3)} l^{24/7}
+f_7^{(4)} l^{32/7},
\end{align} 
and found the best values of the fitting for each value of $\varphi$.
Note that $\tilde{Y}_{a,s}$ and hence $A_{\rm free}$ depend on 
the phases only through $\varphi$ in this case.
In  Fig.~\ref{fig:phi-dep}, 
we plot the $\varphi$-dependence of the coefficients $b$  and $f_7^{(2)}$.
The solid lines represent our analytic prediction while the dots show the numerical 
data from the TBA equations.
Our analytic expressions show a good agreement with the numerical data,
which strongly supports the correspondence to the non-unitary $W$ minimal
models proposed in subsection \ref{subsection:nonuniW}, as well as the continuation
to the complex masses.

\section{UV expansion of T-functions}

To derive the UV expansion of the remainder function,
we need to expand the Y-/T-functions, as well as the free energy part discussed in 
the previous section. This is achieved 
by using an interesting relation between the T-function and the $g$-function 
(boundary entropy) \cite{Bazhanov:1994ft,Dorey:1999cj}. 
Here, we extend the discussion for the minimal surfaces in AdS$_{3}$ 
\cite{Hatsuda:2011ke,Hatsuda:2011jn} to the AdS$_{4}$ case.
We concentrate on the case with $n \notin 4 \bbZ$.

\subsection{T-functions for $\grp{SU}(N)_{4}/\grp{U}(1)^{N-1}$ HSG model}

The first step to derive the expansion of $T_{a,s}$ is to compare the integral equations
for the $T$- and $g$-functions of the $\grp{SU}(N)_{k}/\grp{U}(1)^{N-1}$ HSG model 
with level $k=4$.
In this subsection, we consider those for the T-functions, which are obtained by
a procedure similar to the one from Y-systems to TBA equations.
The extension to general $k$ may be straightforward.
For the reason explained in the next subsection, we also set $m_s$ to be real.

Let us start our discussion by considering the asymptotic behavior 
of $T_{a,s}$ for large $ |\theta |$. To see this, we  note that, 
when $N \in  2\bbZ +1$ with the boundary conditions (\ref{Tbd1}), (\ref{Tbd2}), 
one can invert the relation 
between the Y- and T-functions (\ref{YTrel}) for AdS$_5$,
to express $T_{a,s}$ by $Y_{a,s}$. This is also possible
for $N \in 4 \bbZ + 2$ after imposing the AdS$_4$ condition 
$Y_{1,s} = Y_{3,s}$ and $T_{1,s} = T_{3,s}$. 
In such cases, the asymptotic behavior of $Y_{a,s}$ (\ref{Yasympt}) implies that of $T_{a,s}$,
\eqb\label{Tasympt}
    \log T_{a,s} \to -\nu_{a,s} \cosh \theta \comma
\eqe
for $ 0 \leq a \leq k(=4)$ and $0 \leq s \leq N$,
where constants $\nu_{a,s}$ are related by (\ref{YTrel}) to $m_{a,s} = M_{a,s} L$ as
\eqb
    m_{a,s} = \nu_{a,s+1} + \nu_{k-a,s-1} - \nu_{a+1,s} - \nu_{a-1,s} \period
\eqe
At the boundary $a=0, k$ or $s = 0,N$,  we have
$\nu_{a,s} = 0$.
The above relation together with the mass ratios (\ref{Mratio}) in turn gives
\eqb\label{nuas}
  \nu_{a,s} = \nu_{s}  \sin\bigl( \frac{\pi a}{k}\bigr)/\sin\bigl( \frac{\pi }{k}\bigr) \comma
\eqe
and hence
\eqb
    m_{s} = \bigl( I_{N-1} -2\cos \frac{\pi}{k} \cdot \One \bigr)_{sr} \nu_{r}
    \comma
\eqe
where   
$\One$ is the $(N-1) \times (N-1)$ unit matrix and  $I_{N-1}$ is the incidence matrix 
for $A_{N-1}$. This relation is inverted as
\eqb
   \nu_{s} = (V_{N-1} \cdot J_{N-1}   \cdot V_{N-1})_{sr} m_{r} \comma
\eqe
where 
\eqb\label{VN}
   (V_{N-1})_{rs}
    := \sqrt{\frac{2}{N}} \sin \frac{rs\pi}{N}
    \comma 
\eqe
and $ (J_{N-1})_{rs} := {\delta_{rs}}/{2(\cos \frac{\pi r}{N} - \cos\frac{\pi}{k})} $. 
We have also used $V_{N-1}^{-1} = V_{N-1}$ and 
\eqb\label{diagIN}
   (V_{N-1} \cdot I_{N-1} \cdot  V_{N-1})_{rs} = \delta_{rs} \cdot 2 \cos \frac{\pi r}{N}      
    \period
\eqe

Given the asymptotics (\ref{Tasympt}) , we next subtract the linear terms 
in $l=ML$ from $\log T_{a,s}$ and define
\eqb
    U_{a,s} := \log(T_{a,s} e^{\nu_{a,s}\cosh \theta}) \comma
\eqe
so that $U_{a,s} \to 0 $ for large $|\theta|$.
From  (\ref{nuas}) as well as 
the T-system (\ref{Tsystem}) and the relation between $T_{a,s}$ and $Y_{a,s}$ 
(\ref{YTrel}) with $T_{k-a,s} = T_{a,s}$, we  find that 
\eqb
     \log \frac{T_{a,s}^{+} T_{a,s}^{-}}{T_{a+1,s} T_{a-1,s}}  
     =  U_{a,s}^{+} + U_{a,s}^{-} -U_{a+1,s} -U_{a-1,s}
     = \log(1+ Y_{a,s}) \period
\eqe
Note that terms with $\nu_{a,s}$ cancel each other due to (\ref{nuas}), and that
 the above relation involves the same $s$ only.
 Assuming that $ U_{a,s} $ are analytic in the strip $ -\pi/k  < \im \theta < \pi/k$ 
and vanishing rapidly enough for large $ |\theta| $,   which is expected 
from the relation to $Y_{a,s}$, one can Fourier-transform the above equations.
Further taking into account the boundary conditions on $T_{a,s}$
and using again (\ref{VN}) and (\ref{diagIN}) with $N$ being replaced
by $k$, we obtain
\eqb\label{UY}
   \tilde{U}_{a,s} = 
   - (V_{k-1} \cdot   J'_{k-1} \cdot
   V_{k-1})_{ab} \, \widetilde{\log(1+Y_{b,s})} \comma
\eqe
where tildes stand for the Fourier transform, 
$ \tilde{f}(\omega) = \int d\theta \, e^{i\omega \theta} f(\theta)$, and
$J_{k-1}'$ is given by 
$ (J'_{k-1})_{ab} := {\delta_{ab}}/{2(\cos \frac{\pi a}{k} - \cosh\frac{\pi \omega}{k})} $.
Taking into account $U_{1,s} = U_{3,s} $ for $k=4$, and Fourier-transforming 
back (\ref{UY}), we find the integral equations
of $U_{a,s}$ for $k=4$:
\eqb\label{Ueq}
     U_{a,s} = {\cal K}_{ab} \ast \log(1+Y_{b,s}) \comma
\eqe
where 
\eqb
     {\cal K}(\theta) = \begin{pmatrix}
                          K_{2}(\theta) & K_{1}(\theta) \\
                          2K_{1}(\theta)  & K_{2}(\theta)
                                 \end{pmatrix} \comma
\eqe
and $K_1, K_2$ are given in (\ref{kernels}).

\subsection{$g$-functions for $\grp{SU}(N)_{4}/\grp{U}(1)^{N-1}$ HSG model}

Next, let us consider the $g$-function or boundary entropy 
for the $\grp{SU}(N)_{4}/\grp{U}(1)^{N-1}$ HSG model.
The $g$-function is associated with a boundary, and hence
with a set of corresponding reflection factors in an integrable quantum field theory.
To keep the boundary integrability, we thus set the resonance parameters 
of the HSG model to be
vanishing, so that the bulk S-matrix has the parity invariance up to constant factors
$\eta_{r,s}$ in the S-matrix (\ref{SmatrixHSG}).  
This corresponds to considering real mass parameters $m_{s}$ 
in the TBA equations. The case of the complex $m_{s}$ is discussed later.

The reflection factors are constrained by the conditions from 
the unitarity, crossing-unitarity and  boundary bootstrap
\cite{Fring:1993mp,Ghoshal:1993tm}.
In our case, they read as
\eqb\label{Rcond}
     && R_{a,s} (\theta)  R_{a,s} (-\theta)  =  1  \comma  \nn \\
   &&  R_{a,s} (\theta)   R_{\bar{a},s} (\theta-i\pi)   =  
     S_{aa}^{ss} (2\theta) \comma  \\
    &&  R_{\bar{c},s} (\theta)  =    R_{a,s} (\theta+ i\bar{u}_{ac}^{b}) 
    R_{b,s} (\theta -  i\bar{u}_{bc}^{a}) 
    S_{ab}^{ss}(2\theta + i\bar{u}_{ac}^{b}-i\bar{u}_{bc}^{a})   \period \nn
\eqe
Here, $\bar{u} = \pi - u$ and we have used $\overline{(a,s)} = (\bar{a},s) = (k-a,s)$.
The location of the poles specified by $u^a_{bc}$ 
is the same as that  for the $A_{k-1}$ minimal ATFT. 
Note that the boundary bootstrap equations involve the same label $s$ only.
Given a set of the reflections factors $R_{a,s}$, one can deform it as
$ R'_{a,s} =  R_{a,s}/Z_{a,s} $ \cite{Sasaki:1993xr}, where the deforming 
factors $Z_{a,s}$  need to satisfy
\eqb\label{Zcond}
   && Z_{a,s}(\theta) Z_{a,s} (-\theta) = 1 \comma \nn \\
   && Z_{a,s}(\theta) = Z_{\bar{a},s} (i \pi -\theta) \comma  \\
   &&  Z_{\bar{c},s} (\theta)  =  Z_{a,s} (\theta+ i\bar{u}_{ac}^{b}) 
    Z_{b,s} (\theta- i\bar{u}_{bc}^{a})   \comma \nn
\eqe
in order to maintain the conditions (\ref{Rcond}).

Assuming the existence of the reflection factors corresponding to
the boundary labeled by  the identity 
operator, $R^{\vert \bf 1 \rangle}_{a,s}$, we then consider the deformed 
reflection factors,
\eqb\label{R/Z}
    R_{a,s}^{\vert b,r; C \rangle} 
    =  R_{a,s}^{\vert \One \rangle}/Z_{a,s}^{\vert b,r; C \rangle}
    \comma
\eqe
where 
\eqb\label{Zas}
   Z_{a,s}^{\vert b,r; C \rangle} 
   = \Bigl[ S^{F}_{ab} \Bigl(\theta + \frac{i\pi}{k} C \Bigr)
    S^{F}_{ab} \Bigl(\theta - \frac{i\pi}{k} C \Bigr) \Bigr]^{\delta_{sr}}
   \comma
\eqe
and $S_{ab}^{F}$ is defined in (\ref{SF}). The deforming factors 
$ Z_{a,s}^{\vert b,r; C \rangle} $ are non-trivial only in the case $s=r$, where
they reduce to those for the minimal ATFT \cite{Dorey:2005ak}. This assures that they 
indeed satisfy the conditions (\ref{Zcond}). We also note that when $k=2$, 
the indices $a, b$ take only $1$, and the deforming factors 
of the form (\ref{Zas}) reduce to 
\eqb
   Z_{1,s}^{\vert 1,r;C \rangle} =  \Bigl[ (1+C)_{\theta} (1-C)_{\theta} \Bigr]^{\delta_{sr}}
   \comma
\eqe
which were used to analyze the T-functions for the minimal surfaces in AdS$_{3}$ 
\cite{Hatsuda:2011ke,Hatsuda:2011jn}.

Given a pair of sets of the reflection factors, the $g$-functions associated 
with the corresponding boundaries satisfy  \cite{Dorey:2004xk,Dorey:2005ak}
\eqb\label{geq}
   \log \frac{g_{\vert \alpha \rangle}(l) }{ 
   g_{\vert \beta \rangle}(l)  } \frac{c_{\vert \beta \rangle}}{c_{\vert \alpha \rangle }}
   =  \frac{1}{4}\sum_{\gamma} \int_{{\mathbb R}} d\theta \, \bigl( \phi_{\gamma}^{\vert \alpha \rangle} (\theta)
   - \phi_{\gamma}^{\vert \beta \rangle} (\theta) \bigr) \log\bigl(1+Y_{\gamma} (\theta) \bigr)
   \comma
\eqe
where $c_{\vert \alpha \rangle}$ are certain constants related to the vacuum degeneracy,
and
\eqb
   \phi_{\gamma}^{\vert \alpha \rangle} 
   := \frac{1}{\pi i} \del_{\theta} \log R_{\gamma}^{\vert \alpha \rangle}(\theta) \period
\eqe
When we choose $ R_{a,s}^{\vert b,r; C \rangle}  $ and  $R_{a,s}^{\vert \One \rangle} $
for the pair, the right-hand side of (\ref{geq}) is determined only through the deforming
factors $Z_{a,s}^{\vert b,r; C \rangle}$. By further using the relations 
 $Y_{a,s}(\theta) = Y_{a,s}(-\theta)$ and $Y_{1,s} = Y_{3,s}$
for $k=4$, we find  that
\eqb\label{Geq}
      G_{a,s}(C) := \log \frac{g_{\vert a,s;C \rangle}(l) }{ g_{\vert \One \rangle}(l) }
    \frac{c_{\vert \One \rangle}}{c_{\vert a,s;C \rangle } } 
   =  \Bigl[ {\cal K}_{ab} \ast \log(1+Y_{b,s}) \Bigr] \Bigl(\frac{i \pi}{k} C \Bigr)
   \period
\eqe
By comparing (\ref{Ueq}) and (\ref{Geq}), we  see that
$G_{a,s} (C) = U_{a,s}\bigl( \frac{\pi i}{k} C \bigr)$. Moreover, 
assuming that $c_{\vert \One \rangle}= c_{\vert a,s;C \rangle }$ as in the 
case of AdS$_{3}$, and subtracting the linear terms in $l \propto \nu_{a,s}$
from both sides, we arrive at the relation,
\eqb\label{GT}
    \frac{ {\cal G}^{(0)}_{ \vert a,s;C \rangle } }{ {\cal G}^{(0)}_{ \vert \One \rangle } } 
    = T_{a,s}\Bigl( \frac{\pi i}{k} C \Bigr) \period
\eqe
The ratios of $
    {\cal G}^{(0)}_{ \vert \alpha \rangle }  := 
    \log g_{\vert \alpha \rangle} - f_{\vert \alpha \rangle} l 
$ 
on the left-hand side, with $f_{\vert \alpha \rangle} $ being a constant, 
are the quantities which are directly computed around the UV limit by 
the conformal perturbation theory 
with boundaries \cite{Dorey:1999cj,Dorey:2005ak}. 

\subsection{Expansion of $T_{a,s}$}

Another input for the expansion of the T-functions is their periodicity.
To see this, we first note that, from the Y-system (\ref{Ysystem}) with 
$Y_{1,s} = Y_{3,s}$ and  the boundary conditions 
given in section 2.1, the Y-functions have the quasi-periodicity, 
\eqb
   Y_{a,s}\Bigl( \theta + \frac{n}{k} \pi i\Bigr) = Y_{a,N-s} (\theta) \comma
\eqe
where $n= N+k$ and $k=4$.
Since our T-functions are expressed by the Y-functions for $N \notin 4\bbZ $, they inherit 
the same quasi-periodicity,
\eqb\label{Tquasiperiod}
   T_{a,s}\Bigl( \theta + \frac{n}{k} \pi i\Bigr) = T_{a,N-s} (\theta)
   \qquad (N \notin 4 \bbZ)  \period
\eqe
Taking also into account the structure of the CPT,
we  find that $T_{a,s}$ are expanded as 
\eqb\label{Texp}
   T_{a,s}(\theta) = \sum_{p,q=0}^{\infty} t^{(p,2q)}_{a,s} l^{(1-\Delta) (p+2q)}
    \cosh\Bigl( \frac{k p}{n} \theta \Bigr) \comma
\eqe
with $t_{a,N-s}^{(p,2q)} = (-1)^{p} t_{a,s}^{(p,2q)}$. For lower orders, one can check  
from the T-system (\ref{Tsystem}) that 
the terms $t_{a,s}^{p,q'}$ with $q'$ odd are indeed absent, and that
the first two non-trivial coefficients are $t_{a,s}^{(0,0)}$ 
and $t_{a,s}^{(2,0)}$. The Y-functions also have similar expansion, the coefficients 
of which are related to $t_{a,s}^{(p,2q)}$ by (\ref{YTrel}) with (\ref{13Id}).

To compute these coefficients using the relation to the $g$-functions (\ref{GT}), 
we still need
to find which boundary the reflection factors $ R_{a,s}^{\vert b,r; C \rangle} $ 
correspond to. For this purpose, we recall that
similar reflection factors for the $\grp{SU}(\tiln-2)_{2}/\grp{U}(1)^{\tiln-3}$ HGS model 
in the AdS$_{3}$ case 
corresponded to a boundary labeled  by a fundamental representation of $\alg{su}(\tiln-2)$. 
It is thus expected  that the reflection factors in the present case also correspond to 
a boundary labeled by a definite representation.
Expressing the weight vector by the Dynkin label as
$ \blambda = [\lambda_{1}, \lambda_{2}, \cdots ]$,
we  infer the following correspondence,
\eqb\label{aslambda}
     \vert a,s ; C  \rangle  \ \longleftrightarrow \  \blambda_{(a,s)} \quad
      {\rm with} \quad (\lambda_{(a,s)})_{j} = a \delta_{s}^{j} 
      \period
\eqe

The result of the CPT and (\ref{GT}) then  give  \cite{Dorey:1999cj,Dorey:2005ak}
\eqb\label{tas00}
 t^{(0,0)}_{a,s} = 
   \frac{S^{(k)}_{\blambda_{(a,s)} \bzero }}{S^{(k)}_{\bzero\bzero}} \comma
\eqe
where  $S^{(k)}_{\blambda\bmu}$ is the modular S-matrix
for $ \grp{SU}(N)_{k}$  given by the formula \cite{Gannon:2001py}, 
\eqb\label{modularS}
   S^{(k)}_{\blambda\bmu}\Eqn{=}
(N+k)^{-(N-1)/2}\frac{i^{N(N-1)/2}}{\sqrt{N}}
\exp\left[
\frac{2\pi i}{N(N+k)}
\left(\sum_{j=1}^{N-1} j(\lambda_j+1)\right)
\left(\sum_{j=1}^{N-1} j(\mu_j+1)\right)
\right],\nn\\
&&\times\det\left(
\exp\left[-\frac{2\pi i}{N+k}
\left(\sum_{j=a}^{N-1}(\lambda_j+1)\right)
\left(\sum_{j=b}^{N-1}(\mu_j+1)\right)
\right]
\right)_{1\le a,b\le N} \period
\eqe
For lower $N$, one can check that (\ref{tas00}) indeed solve the constant T-system,
in which $T_{a,s}$ are set to be constants corresponding to the UV limit. This 
provides a justification of 
the correspondence (\ref{aslambda}). 

From  the fact that the HSG model is obtained from an integrable deformation 
of the coset model by weight-zero adjoint operators, the CPT  also gives
\cite{Dorey:1999cj,Dorey:2005ak}
\eqb\label{t20as}
   {t_{a,s}^{(2,0)} \over t_{a,s}^{(0,0)}}
   = - \kappa_{n} G(\tilde{M}_{s}) \cdot \frac{B(1-2\Delta, \Delta)}{2(2\pi)^{1-2\Delta}}
   \left( \frac{S^{(k)}_{\blambda_{(a,s)} \brho_{\rm adj}} }{S^{(k)}_{\blambda_{(a,s)} \bzero} }
   \sqrt{ \frac{ S^{(k)}_{\bzero \bzero} }{ S^{(k)}_{\bzero \brho_{\rm adj} } } }
   - \sqrt{ \frac{  S^{(k)}_{\bzero \brho_{\rm adj} } }{ S^{(k)}_{\bzero \bzero} } }
    \right) \comma
\eqe 
at the next non-trivial order. Here, $\brho_{\rm adj} = [1, 0, ..., 0,1]$ 
and $\bzero = [0, ..., 0]$ are
the  Dynkin labels of the adjoint and the vacuum representation of $\alg{su}(N)$, respectively.

Now, we are in the position to consider the expansion for complex $m_s$.
As discussed in \cite{Hatsuda:2011ke}, the T-functions in this case are expanded as 
\eqb
   T_{a,s}(\theta) = \sum_{p,q=0}^{\infty} \frac{1}{2} 
    \Bigl( t^{(p,2q)}_{a,s} e^{-\frac{k p}{n} \theta}  +   \bar{t}^{(p,2q)}_{a,s} e^{\frac{k p}{n} \theta}
      \Bigr) l^{(1-\Delta) (p+2q)}\comma
\eqe 
where the coefficients  are argued to be continued from the real-mass case
as $ t^{(p,2q)}_{a,s} (m_s) \to t^{(p,2q)}_{a,s} (|m_s|e^{i\varphi_s})$.
We confirm below that the expansions obtained in this way indeed agree
with  numerical results. 

\subsection{Case of $N=2$ ($n=6$)}

In the next section, we discuss the UV expansion of the remainder functions 
for the $6$- and  $7$-cusp minimal surfaces, which correspond to $N=2$ and $N=3$, respectively.
Here, we list the relevant data for the expansion of $T_{a,s}$.

First, when $N=2$, the coefficients $t_{a,s}^{(p,2q)}$ with $p$ odd vanish due to
 (\ref{Tquasiperiod}).
At the lowest order, (\ref{tas00}) and (\ref{modularS}) give%
\footnote{
For $ \grp{SU}(2)_{k}$, the modular S-matrix (\ref{modularS}) simplifies to
$S_{ab}^{(k)} = \sqrt{ \frac{2}{k+2} }\sin\Bigl(\frac{(a+1)(b+1)\pi}{k+2}\Bigr)$.
}
\eqb\label{t00n=6}
    t_{1}^{(0,0)}  = t_{3}^{(0,0)}= \sqrt{3} \comma \qquad t_{2}^{(0,0)} = 2 \comma
\eqe
where we have omitted the index $s$ since it takes 1 only in this case.
Denoting $\blambda_{a,s}$ by $a$ and $\brho_{\rm adj} = \blambda_{2,s}$ by $2$,
the ratios of the modular S-matrix elements appearing in $t_{a}^{(2,0)}$ are
\eqb
    \frac{S^{(4)}_{02}}{S^{(4)}_{00}} =2  \comma  
    \quad 
    \frac{S^{(4)}_{12}}{S^{(4)}_{10}}  = \frac{S^{(4)}_{32}}{S^{(4)}_{30}} =  0 
    \comma \quad
     \frac{S^{(4)}_{22}}{S^{(4)}_{20}}  = -1 \period
\eqe
Collecting these results, we find
\eqb\label{t20n=6}
   t_{1}^{(2,0)} \Eqn{=} 
   \kappa_{6} G \cdot \sqrt{6}\frac{B(\frac{1}{3}, \frac{1}{3})}{2(2\pi)^{1/3}} 
    \comma \qquad 
  t_{2}^{(2,0)} =  \sqrt{3} t_{1}^{(2,0)}    
   \comma
\eqe
where $\kappa_{6} G $ is given by  (\ref{kappa6}) with $G=\tilde{M}_{1} = 1$.
In addition, substituting the expansion (\ref{Texp}) into the T-system (\ref{Tsystem})
with $T_{1,1} = T_{3,1}$, 
we also find that
$  t_{a}^{(0,2)}  = 0 $, $ t_{2}^{(2,0)} = \sqrt{3} t_{1}^{(2,0)}$ and
\eqb\label{t04n=6}
        \begin{pmatrix} 
        t_{1}^{(0,4)} \\  t_{2}^{(0,4)} 
        \end{pmatrix}
         \Eqn{=} 
     \frac{1}{24\sqrt{3}} \left(\begin{array}{cc} 6 & 1 \\ 6\sqrt{3} & 2\sqrt{3} \end{array}\right)
     \begin{pmatrix} (t_{1}^{(2,0)})^{2} \\
       (t_{2}^{(2,0)})^{2} 
       \end{pmatrix}
        \period
\eqe
The ratio of $t_{1}^{(2,0)}$ and $t_{2}^{(2,0)}$ from the T-system agrees with (\ref{t20n=6}),
which provides a non-trivial check of our computations.

In \cite{Hatsuda:2010vr}, the expansion of the Y-functions for $N=2$ was numerically 
determined up to and including ${\cal O}(l^{4/3})$.  
We can compare this with the above results.
To this end, we note that 
the relation between the Y- and T-functions 
in this case reads as $ Y_{1} = 1/T_{2} $, $Y_{2} =1/(T_{1})^{2}$, and that the Y-functions
in \cite{Alday:2009dv,Hatsuda:2010vr} and those in this paper are inverse to each other, 
$Y_{a}^{\rm AGM} = 1/Y_{a}^{\rm here}$.%
\footnote{
We also need to rename the $i^{\rm th}$ cusp  to the $(i+1)^{\rm th}$ cusp to  
match the conventions.
}
Then,
\eqb
   Y_{2}^{\rm AGM} =  (T_{1})^{2} \Eqn{=}  (t_{1}^{(0,0)})^{2} + 2 t_{1}^{(0,0)} t_{1}^{(2,0)}
    l^{4/3} \cosh\Bigl( \frac{4}{3} \theta \Bigr) + {\cal O}( l^{8/3}) \nn \\
     \Eqn{\approx}  3+ 5.4805 |Z|^{4/3}  \cosh\Bigl( \frac{4}{3} \theta \Bigr) + {\cal O}( l^{8/3})
     \comma
\eqe
for real $m_{s}$, 
where $|Z| = l/2$. This agrees with the result in \cite{Hatsuda:2010vr}.

\subsection{Case of $N=3$ ($n=7$)}

When $N=3$,  it follows  from  (\ref{Tquasiperiod}) that $t_{a,2}^{(p,q)} = (-1)^{p}t_{a,1}^{(p,q)}$,
and thus the independent variables are $t_{a,1}^{(p,q)}$ only.
At the lowest order, (\ref{tas00}) and (\ref{modularS}) give
\eqb
    t_{1,1}^{(0,0)} =   t_{3,1}^{(0,0)}  = \frac{\sin\frac{3}{7} \pi}{\sin\frac{\pi}{7}} \comma \qquad
    t_{2,1}^{(0,0)} = \frac{\sin\frac{3}{14} \pi}{\sin\frac{\pi}{14}} \period
\eqe
The ratios of the modular S-matrix elements appearing in $t_{a,s}^{(2,0)}$ are
\eqb
   &&  \frac{S^{(4)}_{\bzero \brho_{\rm adj} }}{S^{(4)}_{\bzero \bzero}}
    = {\sin \frac{5\pi}{14} \over \sin \frac{\pi}{14} }  \comma 
    \quad    \frac{S^{(4)}_{\blambda_{(1,s)} \brho_{\rm adj} }}{S^{(4)}_{\blambda_{(1,s)} \bzero}}
    =  \frac{S^{(4)}_{\blambda_{(3,s)} \brho_{\rm adj} }}{S^{(4)}_{\blambda_{(3,s)} \bzero}}
    = 1  \comma \quad 
      \frac{S^{(4)}_{\blambda_{(2,s)} \brho_{\rm adj} }}{S^{(4)}_{\blambda_{(2,s)} \bzero}}
    =  - {\sin \frac{\pi}{14} \over \sin \frac{3\pi}{14} }  \comma \nn
\eqe
for  $s=1,2$ . Collecting these results gives
\eqb\label{t20n=7}
   t_{1,s}^{(2,0)} = -\kappa_{7} G  \cdot 
  \frac{B(\frac{1}{7},\frac{3}{7})}{2(2\pi)^{1/7}} \frac{\sin \frac{3\pi}{7}}{\sin\frac{\pi}{7}}
   \left( \sqrt{\frac{\sin \frac{\pi}{14}}{\sin \frac{5\pi}{14}}} -
   \sqrt{\frac{\sin \frac{5\pi}{14}}{\sin \frac{\pi}{14}}}  \right)\comma 
   \quad 
   t_{2,s}^{(2,0)} = 2 \cos \frac{\pi}{7} \cdot  t_{1,s}^{(2,0)}
   \comma
\eqe
where $\kappa_{7} G$ is given in (\ref{kappa7G}).
In addition, substituting the expansion (\ref{Texp}) into the T-system (\ref{Tsystem})
with $T_{1,s} = T_{3,s}$, we  find that
$t_{a,s}^{(0,2)} = t_{a,s}^{(1,0)} =0$, $t_{2,s}^{(2,0)} 
= 2\cos \frac{\pi}{7} \cdot  t_{1,s}^{(2,0)}$, and
\eqb\label{t04n=7}
  \begin{pmatrix} 
   t_{1,1}^{(0,4)} \\
   t_{2,1}^{(0,4)} 
   \end{pmatrix}  \Eqn{=} 
 \frac{s_0}{2(1+s_0)(3-4s_0)}
\begin{pmatrix} 2+s_0-4s_0^2 & s_0 \\
  s_0^{-1} & 1-s_0 
  \end{pmatrix}
  \begin{pmatrix}
    (t_{1,1}^{(2,0)})^{2} \\
      (t_{2,1}^{(2,0)})^{2} 
    \end{pmatrix}   \comma
\eqe
where $s_{0} = \sin(\pi/14)$.  
The ratio of $t_{1,s}^{(2,0)}$ and $t_{2,s}^{(2,0)}$ from the T-system 
agrees with (\ref{t20n=7}),
which provides a non-trivial check of our computations again.

\begin{figure}[t]
 \begin{minipage}{0.49\hsize}
  \begin{center}
   \includegraphics[width=80mm]{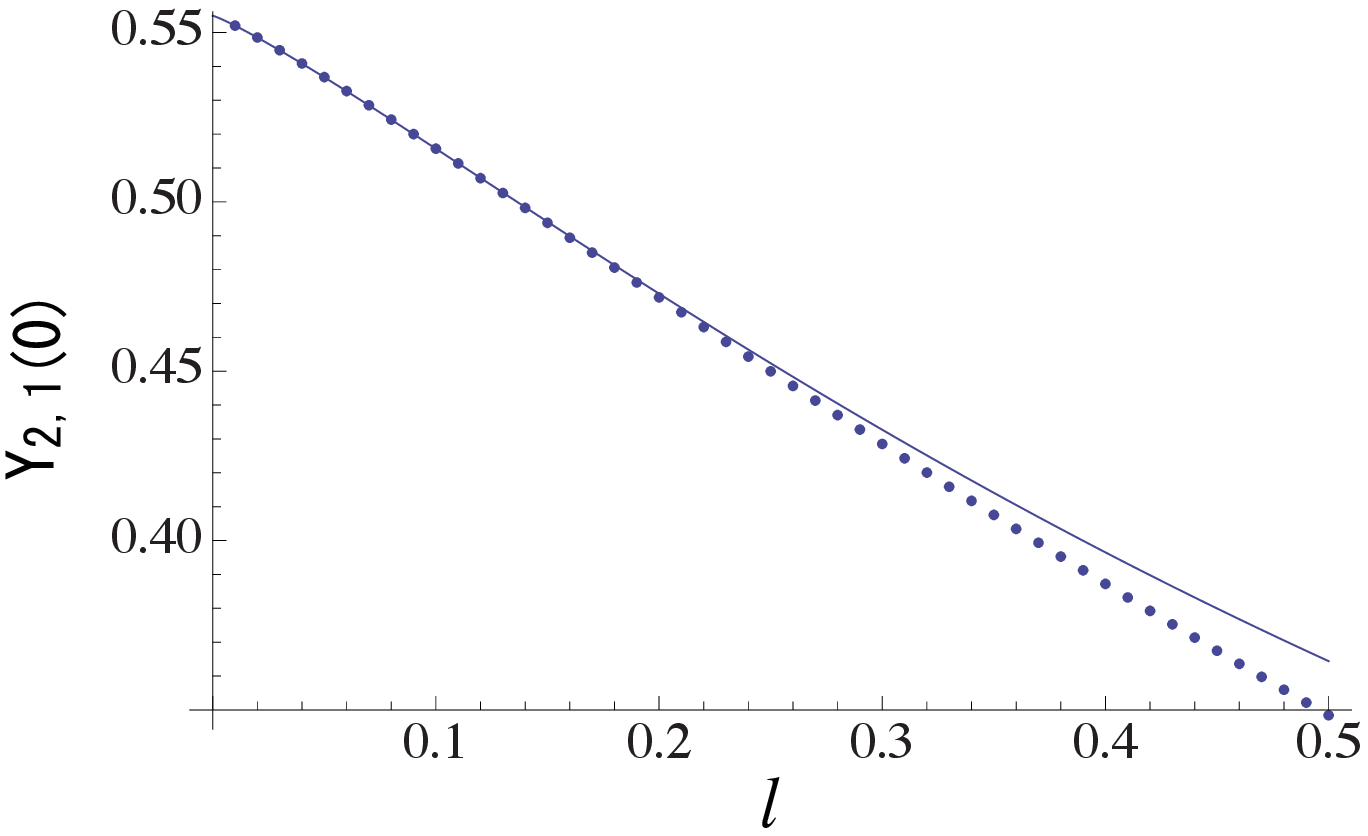} \\ (a)
  \end{center}
 \end{minipage}
 \begin{minipage}{0.49\hsize}
  \begin{center}
   \includegraphics[width=80mm]{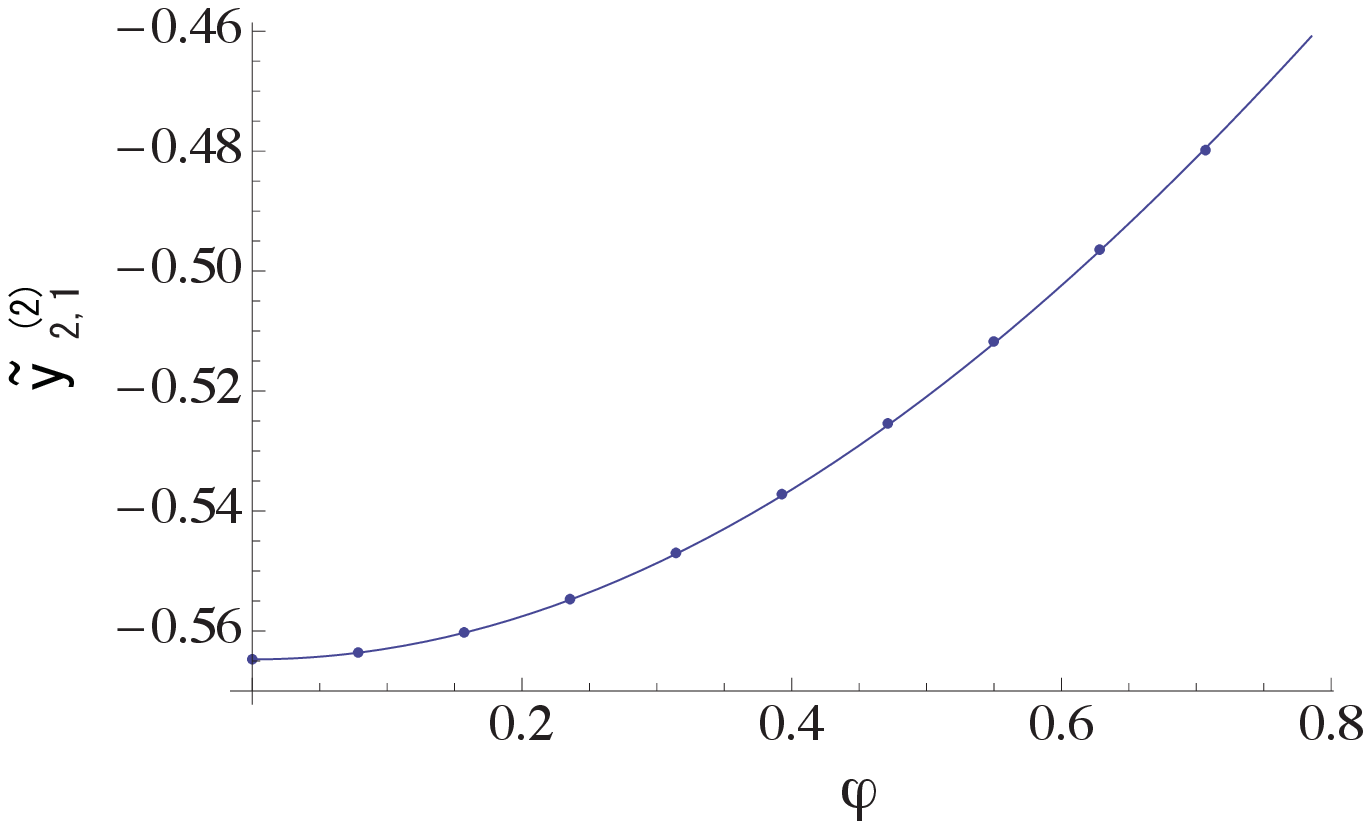} \\ (b)
  \end{center}
 \end{minipage}
 \caption{(a) Plots of $Y_{2,1}(0)$ from numerics (points) and from the analytic
 expansion (solid line) for $m_{1} = m_{2} = l$.
\  (b) Phase dependence of an expansion coefficient of $\tilde{Y}_{2,1}(0)$. The points
are from numerical fitting, whereas  the solid line represents the analytic 
expression.}
\label{fig:y21}
\end{figure}

We have checked our analytic expansion by comparing it with numerical results.
For example, Fig.~\ref{fig:y21} (a) 
shows plots of $Y_{2,1}(0)$ from numerics (points) and from our expansion (solid line)
for real and equal mass parameters $m_{1} = m_{2} = l$, which are in good agreement
with each other around the UV limit. To check the phase dependence, we have also 
numerically solved
the TBA equations  from $l=1/20$ to $3/2$ for $m_{s} = e^{i\varphi_{s}} l$ with 
$\varphi= \varphi_{1} - \varphi_{2} = -\pi j/40 $ $(j=0, ..., 9)$, and fitted 
$\tilde{Y}_{2,1} (0)$ by the function
\eqb
    \tilde{Y}^{\rm (fit)}_{2,1}(0) = \tilde{Y}^{(0)}_{2,1} + \sum_{p=1}^{10}  
    \tilde{y}_{2,1}^{(p)}(\varphi) l^{4p/7} \period
\eqe
Here, $\tilde{Y}^{(0)}_{2,1} = t_{2,1}^{(0,0)}/(t_{1,1}^{(0,0)})^{2} \approx 0.554958$
is the exact value in the UV limit.
The points in Fig.~\ref{fig:y21} (b) show the fitted values of 
$ \tilde{y}_{2,1}^{(2)} $ for each $\varphi$. 
We find a good agreement with our analytic expression (solid line) again.

\section{UV expansion of remainder function}\label{sec:UV-rem}

Based on the results so far, we derive the UV expansion 
of the remainder function in this section. 

\subsection{Remainder function for six-cusp minimal surfaces}

In the case of $n=6$ ($N=2$), the relevant cross-ratios for $\Delta A_{\rm BDS}$
in (\ref{delABDSn=6}) are 
\eqb\label{crn=6}
    u_{r,r+3}  = \bigl( U_{1}^{[2r-1]} \bigr)^{-1} =  \frac{1}{T_{2,1}^{[2r-2]}T_{2,1}^{[2r]}} \period
\eqe
These are also rewritten by using  $T_{2,1}^{[6+r]} = T_{2,1}^{[r]}$.
In the UV limit, the cross-ratios become $u_{r,r+3} = 1/4$ and equal to each other.
Form (\ref{crn=6}) and the expansion of the T-functions, one finds that $\Delta A_{\rm BDS}$
is expanded in terms of $t_{a,s}^{(0,0)}$, $(t_{a,s}^{(2,0)})^{2}$ and $t_{a,s}^{(0,4)}$ 
for real $m_{s}$ up to ${\cal O}(l^{4(1-\Delta)})$. Similarly to the AdS$_{3}$ case 
\cite{Hatsuda:2011ke,Hatsuda:2011jn}, other $t_{a,s}^{(p,2q)}$ do not appear in the 
expansion due to the $\bbZ_{n}$-symmetry. Further using (\ref{t20n=6}) and (\ref{t04n=6}),
we find 
\eqb
\Delta A_{\rm BDS}=
   \frac{3}{4} \dilog(-3) 
   - \frac{1}{16}(3+2\log 2) (t_{2,1}^{(2,0)})^{2} l^{4(1-\Delta)}+\cO(l^{6(1-\Delta)})
\period
\eqe
Since the period term and the bulk term in the free energy part cancel each other,
we arrive at the expansion of the remainder function,
\eqb\label{R6}
   R_6 = \frac{\pi}{6} +  \frac{3}{4} \dilog(-3) + \Bigl[ \frac{\pi}{6} 
       C_{6}^{(2)} - \frac{1}{16}(3+2\log 2) (\tilde{t}_{1,1}^{(2,0)})^{2} \Bigr] 
         (\kappa_{6}G)^{2}\cdot 
          l^{8/3}+\cO(l^{4}) \comma
\eqe
where we have introduced 
\eqb
      \tilde{t}_{a,s}^{(2,0)} := {t}_{a,s}^{(2,0)}/\kappa_{n} G \comma
\eqe
and      
$C_{6}^{(2)}$, $t_{1,1}^{(2,0)}$ and $\kappa_6 G$ are given 
by (\ref{eq:C_n^(2)}), (\ref{t20n=6}) and 
(\ref{kappa6}) with $G= \tilde{M}_{1} =1$, respectively.
For complex $m_{s}$, one has only to replace $G^{2}(\tilde{M}_{s})$ by 
$| G(\tilde{M}_{s}e^{i\varphi_{s}}) |^{2}$, giving just $|G|^2 =1$.
These results agree with those in \cite{Hatsuda:2010vr}.

\subsection{Remainder function for seven-cusp minimal surfaces}

In the case of $n=7$ ($N=3$), the relevant cross-ratios for $\Delta A_{\rm BDS}$
in  (\ref{delABDSn=7}) are 
\eqb\label{crn=7}
   u_{r,r+3}  = \bigl( U_{2}^{[2r-2]} \bigr)^{-1} 
   = \frac{T_{2,1}^{[2r-2]}}{T_{2,2}^{[2r-3]} T_{2,2}^{[2r-1]}}
   \period
\eqe
These are also rewritten by using  $T_{2,1}^{[7+r]} = T_{2,2}^{[r]}$.
From (\ref{crn=7}) and the expansion of the T-functions, one finds again 
that $\Delta A_{\rm BDS}$
is expanded in terms of $t_{a,s}^{(0,0)}$, $(t_{a,s}^{(2,0)})^{2}$ and $t_{a,s}^{(0,4)}$ 
for real $m_{s}$ up to ${\cal O}(l^{4(1-\Delta)})$. 
Further using (\ref{t20n=7}) and (\ref{t04n=7}), we  find that 
\eqb
\Delta A_{\rm BDS}=D_{7}^{(0)}+D_{7}^{(4)} (t_{1,1}^{(2,0)})^2
 l^{4(1-\Delta)}+\cO(l^{6(1-\Delta)})
\comma 
\eqe
where
\eqb
D_{7}^{(0)} \Eqn{=}-\frac{7}{4}\left[ \log^2 \( 2\cos \frac{\pi}{7} +1\)
+\Li_2\(\frac{2\cos \frac{\pi}{7}}{2\cos \frac{\pi}{7}+1}\) \right] \comma  \\
D_{7}^{(4)} \Eqn{=}-\frac{7}{16s_0(15+5s_0-24s_0^2)}
\bigl[ s_0(3-4s_0)+(1-s_0)(1-4s_0) \log (3-4s_0^2) \bigr] \period \nn
\eqe

Due to the cancelation between the period and bulk terms, 
we arrive at the expansion of the remainder function,
\eqb\label{R7}
    R_7 = \frac{\pi}{6} c_{7}  + D_{7}^{(0)} 
    + \Bigl[ \frac{\pi}{6}C_{7}^{(2)}  + D_{7}^{(4)} (\tilde{t}_{1,1}^{(2,0)})^{2}\Bigr]  
    (\kappa_{7} G)^{2}
    \cdot l^{16/7}+\cO(l^{24/7}) \comma
\eqe
where $C_{7}^{(2)}$, $t_{1,1}^{(2,0)}$ and $\kappa_{7} G(\tilde{M}_{s})$ 
are given by (\ref{eq:C_n^(2)}), (\ref{t20n=7}) and
(\ref{kappa7G}), respectively.
For complex $m_{s}$, $ \bigl(\kappa_{7}G(\tilde{M}_{s}) \bigr)^{2}$ is replaced by 
$| \kappa_{7}G(\tilde{M}_{s}e^{i\varphi_{s}}) |^{2}$.

\begin{figure}[t]
 \begin{center}
  \begin{minipage}{0.5\hsize}
  \begin{center}
   \includegraphics[width=80mm]{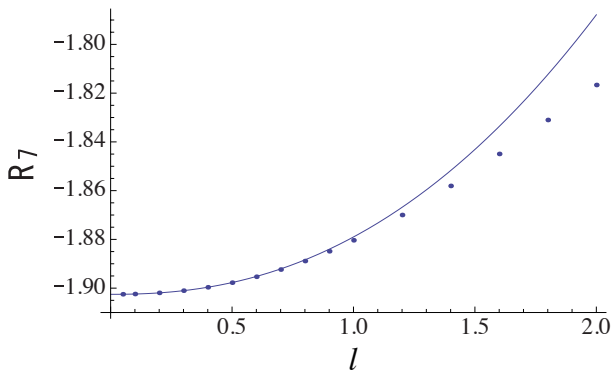}
  \end{center}
 \end{minipage}
 \caption{7-point remainder function at strong coupling. The points are 
 from numerics, whereas the solid line is from our analytic expansion (\ref{R7}). These are
 evaluated for $m_{1} = e^{-\frac{\pi}{40} i} l $, $m_{2} = e^{-\frac{\pi}{20} i} l $.}
  \label{fig:R7}
  \end{center}
\end{figure}

In Fig.~\ref{fig:R7}, we show plots of the 7-point (7-cusp) 
remainder function 
for $m_{1} = e^{-\frac{\pi}{40} i} l $, $m_{2} = e^{-\frac{\pi}{20} i} l $
from numerics (points)
and from our analytic expansion (solid line). 
They are in good agreement around the UV limit.

\subsection{Rescaled remainder function}

In \cite{Brandhuber:2009da}, it was observed  numerically for the 8-cusp 
minimal surfaces in AdS$_{3}$ that the remainder functions at strong coupling
and at two loops are close to each other, but different, if they are appropriately
shifted and rescaled. In \cite{Hatsuda:2011ke,Hatsuda:2011jn}, 
this was analytically demonstrated around the UV limit  
for the general null-polygonal minimal surfaces
in AdS$_{3}$. 

Similarly, one can define the rescaled remainder function for the AdS$_{4}$ case by
\eqb\label{rescaledR}
   \bar{R}_{n} := \frac{R_{n} - R_{n,{\rm UV}}}{R_{n,{\rm UV}} - R_{n,{\rm IR}}} \period
\eqe 
Here, $R_{n,{\rm UV}}$ is the $n$-point remainder function 
in the UV limit, which is read off from (\ref{R6}) and (\ref{R7}).
 $R_{n,{\rm IR}}$ is the $n$-point remainder function 
in the IR limit where $|m_{s}| \to \infty$. To find this constant, we  note 
the asymptotics of $Y_{a,s}$ (\ref{Yasympt}) valid for real $m_{s}$
and $ | \im \theta \, | < \pi/2$, and successively use the Y-system (\ref{Ysystem}),
to express $Y_{a,s}^{[r]}$, e.g., by $Y_{a,s}^{[0]}$ and $Y_{a,s}^{[-1]}$ in the IR limit. 
We then find that
\eqb
    (u_{1,4},  u_{2,5}, u_{3,6}) \sim (e^{-m_{1}}, 1,e^{-m_{1}}) \comma
\eqe
for $n=6$ \cite{Alday:2009dv}, and that 
\eqb
    (u_{1,4},  u_{2,5}, u_{3,6}, u_{4,7},  u_{1,5}, u_{2,6},u_{3,7} )
    \sim \Bigl(e^{-\sqrt{2} m_{2}}, \frac{1}{2}, 1, e^{-m_{1}}, e^{-m_{1}}, 1,\frac{1}{2} \Bigr)
    \comma
\eqe
for $n=7$. 
The leading terms in $\Delta A_{\rm BDS}$ thus cancel $A_{\rm periods}$
\cite{Yang:2010as}. Furthermore, from the fact that $A_{\rm free}$ vanishes in the IR limit, 
it follows that
\eqb
   R_{6,{\rm IR}} = -\frac{\pi^2}{12} \comma \qquad
    R_{7,{\rm IR}} = -\frac{\pi^2}{6} \period
\eqe
Substituting these values into (\ref{rescaledR}),   we find that 
\eqb\label{Rbarn=6}
   \bar{R}_{6} \approx -0.0528126 l^{8/3} + {\cal O}(l^{12/3}) \comma
\eqe
for $n=6$, and
\eqb
   \bar{R}_{7} \approx -0.707647 |\kappa_{7} G|^{2} l^{16/7} + {\cal O}(l^{24/7}) \comma
\eqe
for $n=7$. 

\begin{figure}[t]
 \begin{minipage}{0.49\hsize}
  \begin{center}
   \includegraphics[width=80mm]{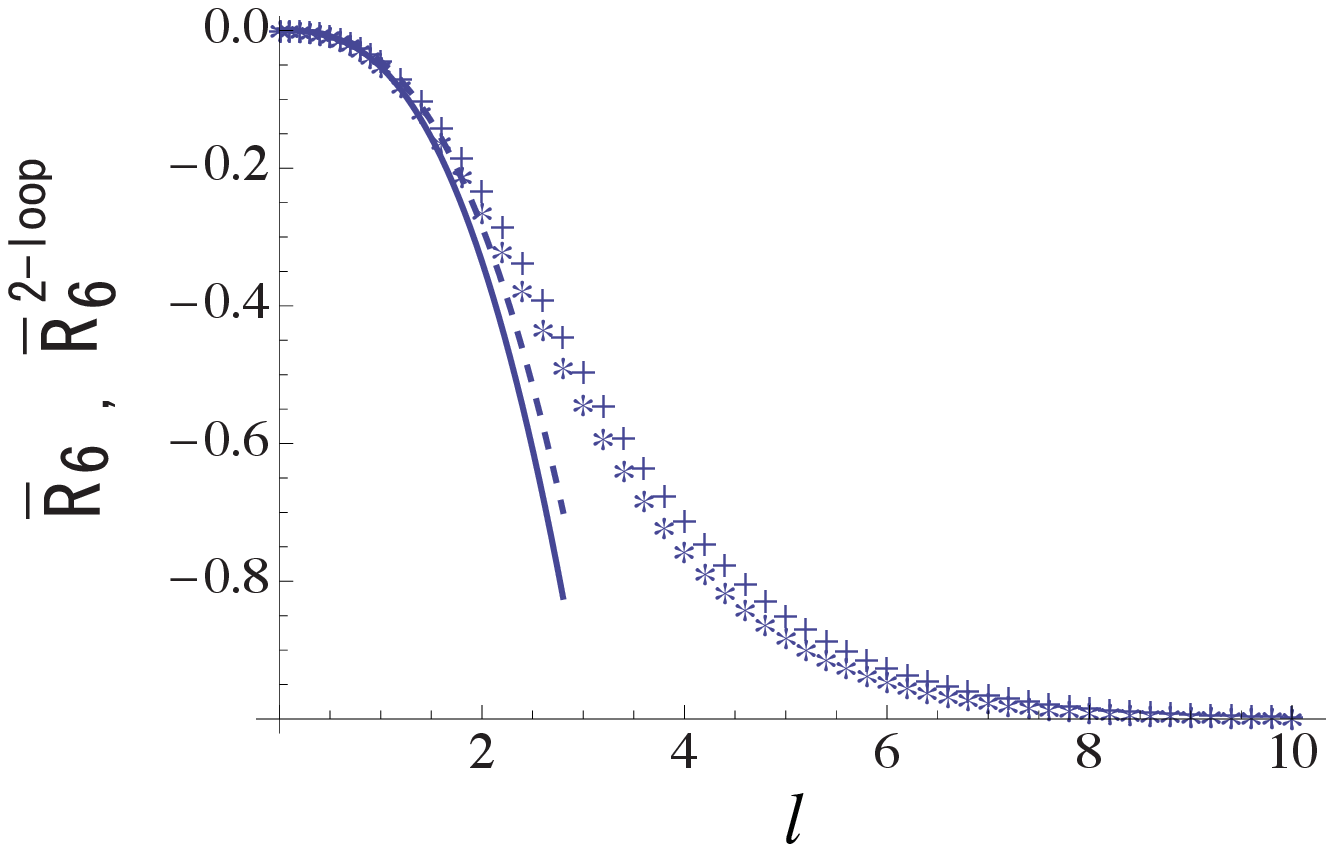} \\ (a)
  \end{center}
 \end{minipage}
 \hspace{-5mm}
 \begin{minipage}{0.49\hsize}
  \begin{center}
  \vspace{-1mm}
   \includegraphics[width=80mm]{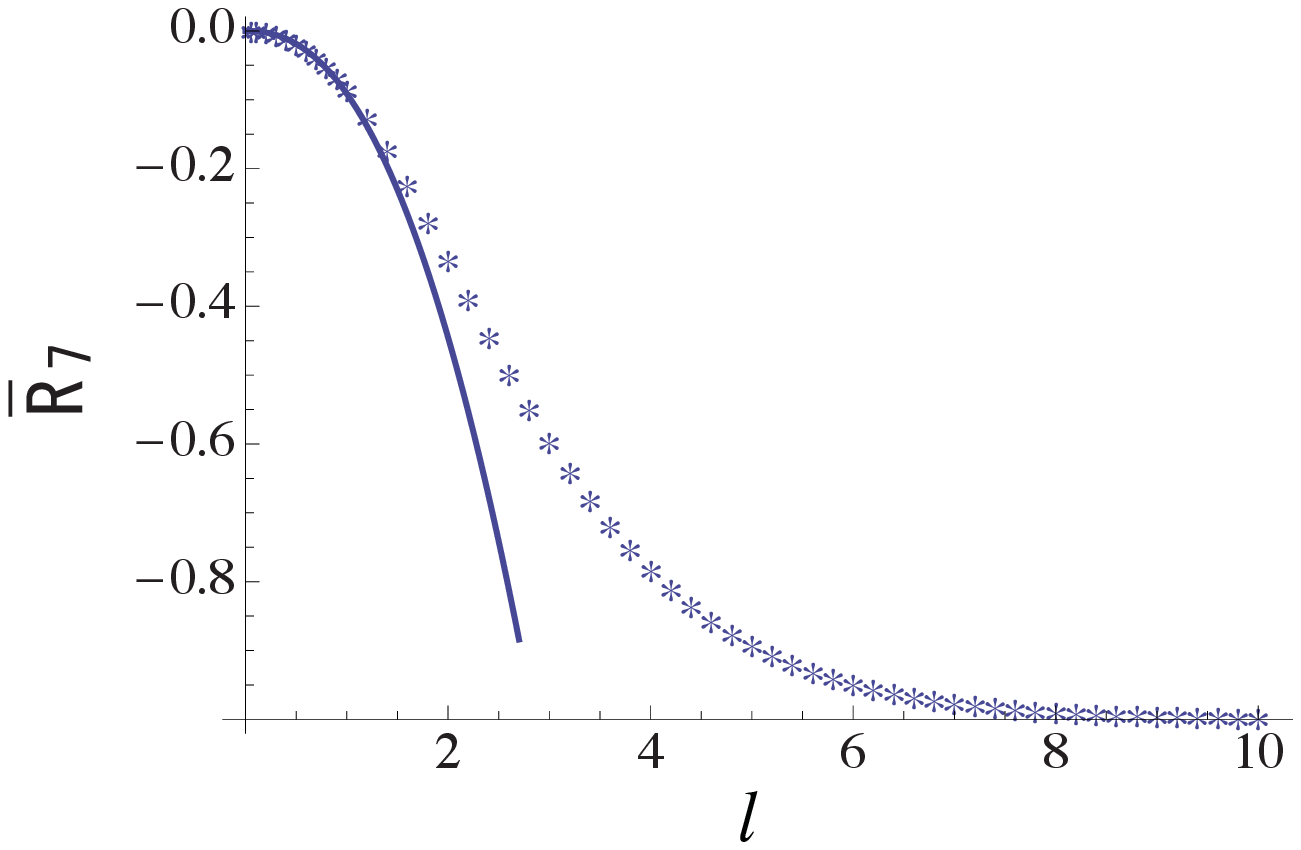} \\ (b)
  \end{center}
 \end{minipage}
 \caption{(a)  6-point rescaled remainder functions. Points denoted by $\ast$/$+$
stand for a plot from numerics at strong coupling/two loops.
The solid/dashed line represents the analytic expansion around the UV limit
at strong coupling/two loops. These are evaluated for $m_{1} = e^{-\frac{\pi}{20} i} l $.
\  (b)  7-point rescaled remainder function at strong coupling. Points are from numerics
and the solid line represents the analytic expansion around the UV limit. 
These are evaluated for $m_{1} = e^{-\frac{\pi}{40} i} l $, $m_{2} = e^{-\frac{\pi}{20} i} l $.}
\label{fig:Rbar}
\end{figure}

On the weak-coupling side, the remainder function at two loops for $n=6$ 
in the AdS$_{4}$ case 
is read off from  the results in the AdS$_{5}$ case 
\cite{Anastasiou:2009kna,DelDuca:2009au,DelDuca:2010zg,Goncharov:2010jf}.
In particular, one can find the UV expansion of the remainder function 
from  a very concise expression in \cite{Goncharov:2010jf}
and the expansion of the T-functions in the previous section:
$ R_{6}^{\rm 2\mbox{-}loop} \approx 1.08917 - 0.0487985 l^{8/3}$. The value in the
UV limit $l \to 0$ has been given in \cite{Anastasiou:2009kna,DelDuca:2010zg}.
The rescaled remainder function is defined similarly to  (\ref{rescaledR}).
Since the two-loop remainder function vanishes in the IR limit,
the rescaled remainder function at two loops is expanded as
\eqb
   \bar{R}_{6}^{\rm 2\mbox{-}loop} \approx -0.0448036 l^{8/3} + {\cal O}(l^{12/3}) \period
\eqe
The ratio of the rescaled remainder functions at strong coupling and at two loops
is then
\eqb
    {  \bar{R}_{6} \over \bar{R}_{6}^{\rm 2\mbox{-}loop}} \approx 1.17876 \comma
\eqe
which is close to 1. By numerics, we also find that the two 6-point rescaled remainder 
functions are close to each other for all the scales as shown in Fig.~\ref{fig:Rbar} (a).
We also show the 7-point rescaled remainder function from the numerics 
in Fig.~\ref{fig:Rbar}  (b).
For both the 6- and 7-point cases, we find a good agreement with our analytic 
expansions around the UV limit.
It would be of  interest to compare the 7-point rescaled remainder function
at strong coupling with the one  at weak coupling, which is yet to be computed.

\subsection{Cross-ratios and mass parameters}

We have expanded the remainder function by the mass parameters.
In order to express it by the cross-ratios, one needs to invert the relation 
between the former and the latter.

For $n=6$, it follows from (\ref{crn=6}) and the expansion of the T-functions
in the previous section that 
\eqb
   u_{r,3+r} = \frac{1}{4} - \frac{1}{8}\cos\Bigl( \frac{1}{3}(4\varphi-(2r-1)\pi)\Bigr) \cdot
      |t_{2}^{(2,0)}| l^{2(1-\Delta)} + \cO(l^{4(1-\Delta)}) \period
\eqe
Inverting this relation, one can express the mass parameter 
$m_1= e^{ i\varphi} l$ by the cross-ratios \cite{Hatsuda:2010vr}. 
In the notation in this paper,  the result reads as
\eqb
   \tan \frac{4}{3}\varphi \Eqn{=} 
      \frac{\sqrt{3}(U_{1}^{[-1]} -U_{1}^{[1]})}{2U_{1}^{[3]}-U_{1}^{[-1]} -U_{1}^{[1]} } \comma
      \qquad 
       l^{\frac{4}{3}} = \frac{-2U_{1}^{[3]}+U_{1}^{[-1]} +U_{1}^{[1]} }{
       6\sqrt{3} |t_{1}^{(2,0)}| \cos \frac{4}{3}\varphi} \period    
\eqe

For $n=7$, it follows from (\ref{crn=7}) and the expansion of the T-function
in the previous section that 
\eqb
   u_{r,r+3} = \frac{1}{t_{2,1}^{(0,0)}} - 
   \frac{2\cos\frac{2\pi}{7}-1}{2(t_{2,1}^{(0,0)})^{2}}
   \Bigl( t_{2,1}^{(2,0)} e^{-\frac{4}{7}\pi(r-1) i} 
   + \bar{t}_{2,1}^{(2,0)} e^{\frac{4}{7}\pi(r-1) i}\Bigr) 
     l^{2(1-\Delta)} + \cO(l^{3(1-\Delta)}) \period
\eqe
By inverting this relation, one can express the mass parameters by the cross-ratios.
For example, when $\varphi_{1} = \varphi_{2}$, the inversion is simple, but generically it is
not.

\section{Conclusions and discussion}

In this paper we have evaluated the regularized area 
of the null-polygonal minimal surfaces in AdS$_4$, and the 
remainder function for the corresponding Wilson loops/amplitudes
at strong coupling. They are described by the TBA integral equations 
or the associated T-/Y-system of the HSG model, which is regarded as 
the integrable perturbation of the generalized parafermion CFT
by the weight-zero adjoint fields.  
The connection to the HSG model
as well as to the corresponding CFT allows us to derive the analytic expansion of 
the remainder function around the UV/regular-polygonal limit
by using the conformal perturbation theory. Generalizing the results in the AdS$_3$ case,
we have found or argued  that  the TBA systems
in the single-mass cases are given by  those for the perturbed $\grp{SU}(4)$ diagonal coset
models and $W$ minimal models.  This is used 
to find the precise expansion coefficients through their mass-coupling relations 
and correlation functions. 
We  have  derived the leading-order expansion explicitly
for $n=6$ and 7. For the 6-point case, we have also compared the
rescaled remainder function with the two-loop one. They 
are close to each other, but different, similarly to the AdS$_{3}$ case.
Although we have focused on the $n \notin 4 \bbZ$ case in this paper,
it would be an interesting problem to generalize our analysis to 
the minimal surfaces with general $n$, and to compare their
remainder functions with those at weak coupling.

As noted in section \ref{section:UVfree}, 
the TBA equations for the AdS$_{4}$ minimal surfaces
generally exhibit a numerical instability around the UV limit.
In spite of that, our analytic expansion works well, which proves our formalism to
be useful in this respect as well. 
It would also be desirable to establish the proposed connection  
to the TBA systems of the non-unitary diagonal 
coset/$W$ minimal models, and to substantiate the ``decomposition" discussed
in section \ref{section:singlemass}.

The remainder function at strong coupling for general
kinematical configurations are given by the minimal surfaces in
AdS$_{5}$. The corresponding TBA system is recovered  by
reintroducing the parameters dropped in the AdS$_{4}$ case. 
For example, in the 6-point case,  the relevant 
chemical potential is turned on by a twist operator in 
the $\bbZ_4$-parafermion or the  $\grp{SU}(2)_4/\grp{U}(1)$ coset  
CFT. It would be interesting to find corresponding  operators for higher-point cases,
as well as to identify the relevant  integrable system and the CFT. 
This would also provide a way to analyze the AdS$_4$ minimal surfaces with 
the chemical potential $\mu = -1$,
which are not discussed in this paper. Finally, it would be very interesting to find 
the quantum/strong-coupling corrections to the minimal surfaces, and possibly 
to extrapolate the results to the weak-coupling side.

\subsection*{Acknowledgements} 
We would like to thank Y.~Aisaka, Z.~Bajnok, J.~Balog, P.~Dorey, C.~Dunning, D.~Fioravanti, 
K.~Hosomichi, J.L.~Miramontes, M.~Rossi, 
K.~Sakai,  V.~Schomerus, J.~Suzuki and  R.~Tateo for useful discussions and conversations. 
We are also grateful to Yukawa Institute for Theoretical Physics, 
where part of this  work was done during  YITP workshop YITP-W-12-05.
Y.~S. would like to thank the organizers of a conference
``Progress in Quantum Field Theory and String Theory" at Osaka,
CQUeST-IEU Focus program at Seoul, APCTP-CQueST-IEU Workshop 
at Jeju Island, Yukawa International Seminar (YKIS) 2012  
at Kyoto, and YIPQS long-term workshop at Kyoto, 
for warm hospitality. The work of Y.~H. is supported in part by the JSPS Research 
Fellowship for Young Scientists, whereas the work of K.~I. and Y.~S. is supported 
in part by Grant-in-Aid for Scientific Research from the Japan Ministry of Education, 
Culture, Sports, Science and Technology.

\appendix
\section{Three-point function in  $W$ minimal models}\label{sec:3-point}

In this appendix we review the free field representation of 
the $WA_{k-1}^{(p,q)}$ minimal
model and compute the three-point function 
of the ground-state and perturbing operators
for $k=4$ and $p=5,q=7$, 
that is used in section \ref{section:UVfree} to analyze the 7-point
remainder function.
To lighten the notation, we refrain from using boldface letters for the weight vectors.

\subsection{Free field representation}

The $WA_{k-1}$ minimal model  \cite{Fateev:1987zh} is realized by 
the scalar fields  $\varphi = (\varphi_1, ..., \varphi_{k-1})$ in 
the $\alg{sl}(k)$ conformal Toda field theory with the Lagrangian,
\eqb
   L = \frac{1}{8\pi} (\del \varphi)^{2} + \tilde{\mu} \sum_{j=1}^{k-1} e^{b(e_{j},\varphi)}
   \period
\eqe
 Here, $e_{j}$ are the simple roots of
$\alg{sl}(k)$, $( \ , \ )$ denotes the inner-product, $\tilde{\mu}$ is 
the scale parameter and $b$ is the 
dimensionless coupling. The system has the background charge,
\eqb
   Q = \Bigl(b+\frac{1}{b} \Bigr) \rho  \comma
\eqe
where $\rho = \sum_{j} \omega_{j}$ is the Weyl vector of $\alg{sl}(k)$ and $\omega_j$ are the 
fundamental weights satisfying $(e_i,\omega_j)=\delta_{ij}$. 
The energy momentum tensor is 
\eqb
   T(z) = -\frac{1}{2} (\del \varphi)^{2} + (Q, \del^{2} \varphi) \period
\eqe
The central charge is given by
\eqb\label{cTFT}
    c = k-1 + 12 Q^{2}  = (k-1) \Bigl[ 1+ k(k+1) \Bigl(b+\frac{1}{b} \Bigr)^{2} \Bigr] \period
\eqe
For the $WA_{k-1}^{(p,q)}$ minimal model  with the central charge (\ref{cWA}), 
the coupling is given by
\eqb
    ib =  \sqrt{\frac{p}{q}}.
\eqe
The primary field of the CFT is represented by the vertex operator, 
\eqb\label{VertexOp}
   V_{\alpha} := e^{(\alpha, \varphi)} \comma
\eqe
which has the dimension
\eqb
   h(\alpha) = h(2Q-\alpha) = \frac{1}{2} (\alpha, 2Q-\alpha) \period
\eqe
Thus, the field with  the dimension  (\ref{h+-}) is represented 
by  the vertex operator with 
\eqb\label{alphaLambda}
   \alpha = \alpha(\Lambda_{+}, \Lambda_{-}) 
     := -\frac{1}{b} \Lambda_{+} - b \Lambda_{-}
     \period
\eqe

For example,
from (\ref{w11ad}), we find that the operator $\phi_{(1,1,{\rm ad})}$ 
corresponds to $V_{\alpha}$ with
\eqb
    \alpha = \alpha_{(1,1,{\rm adj})} :=  -b e_{0} \comma
\eqe
where 
\eqb
     e_{0} = \sum_{j=1}^{k-1} e_{j} = \omega_{1} + \omega_{k-1} 
\eqe
is the highest root, i.e., the highest weight of the adjoint representation.
Similarly, from (\ref{gs}),  the ground-state operator for 
$k=4$ and 
$p=n-2, q=n$  with $n$ odd corresponds to
\eqb
  \alpha = \alpha_{0} := -\Bigl( \frac{n-7}{2} b^{-1} + \frac{n-5}{2} b \Bigr) 
  \omega_{2} \period
\eqe
For $n=7$, this simplifies to
\eqb
  \alpha_{0} = -b \omega_{2}.
\eqe

\subsection{Three-point function}

We are interested in the three-point structure constant 
$C_{\phi_0  \phi_{(1,1, {\rm adj})}  \phi_0} $ of the 
ground-state operator $\phi_0$ and the perturbing operator
$\phi_{(1,1,{\rm adj})} $
for $k=4$ and $p=5, q=7$, 
where
\eqb
  \phi_{0} \sim  V_{\alpha_{0}} \comma \quad  \phi_{(1,1, {\rm adj})}  \sim  V_{-b e_{0}} \comma
\eqe
up to normalization. 
To compute this, we first note that 
the normalized three-point function is
generally given by \cite{Fateev:2007ab,Chang:2011vka} 
\eqb\label{eq:norm3pt}
   \big\bra \calO_{\alpha_{1}} \calO_{\alpha_{2}} \calO_{\alpha_{3}}\big
   \ket_{\rm normalized}
   = B(\alpha_{1}) B(\alpha_{2}) B^{-1}(\alpha_{3}) 
   \big\bra V_{\alpha_{1}} V_{\alpha_{2}} V_{2Q -\alpha^{\ast}_{3}}\big
   \ket \comma 
\eqe
where ${\cal O}_\alpha := B(\alpha) V_\alpha$ is the normalized operator 
so that $\bigl\bra {\cal O}_\alpha (z) {\cal O}_{\alpha^\ast} (0) \bigr\ket_{\rm normalized} 
= |z|^{-4h(\alpha)} $,
and  $\alpha^{\ast}$ for $\alpha$ is defined 
through $(\alpha,e_{j}) = (\alpha^{\ast},e_{k-j})$.
The normalization constant $B(\alpha)$ is given by 
\eqb
 B(\alpha) \Eqn{=} \sqrt{A(\alpha) A(2Q) \over A(2Q-\alpha) A(0)} \comma \nn \\
 A(\alpha) \Eqn{=} \Bigl( \pi \tilde{\mu} \gamma(b^{2}) \Bigr)^{(\alpha-Q,\rho)/b} 
 \prod_{ e > 0}
 \Gamma\bigl(1-b(\alpha-Q,e) \bigr) \Gamma\bigl(-b^{-1}(\alpha-Q,e) \bigr) \comma
\eqe
with the product being over the positive roots. The normalization of the vacuum has also 
been taken into account.
Next, the unnormalized structure constant  $C^{\alpha_3}_{\alpha_1,\alpha_2}$ 
is obtained by extracting the residue of the Coulomb-gas integral for 
$\bigl\bra V_{\alpha_1} (\infty) V_{\alpha_2} (1) V_{2Q-\alpha_3} (0) \bigr\ket$.
In particular, a class of the structure constants 
has been evaluated in eq. (1.53) 
of \cite{Fateev:2007ab}: 
\eqb\label{defC}
   C^{\alpha}_{-be_{0}, \alpha} 
   \Eqn{=}
   \sum_{i=1}^{k} \prod_{j\neq i}^{k}
   { \pi \tilde{\mu} \gamma\bigl( b(\alpha -Q, h_{ji}) \bigr)\over
   \gamma(-b^{2}) \gamma\bigl( 1+ b^{2} + b(\alpha -Q, h_{ji}) \bigr) }
   \calF_{i}^{2}(\alpha)  \comma
\eqe
where 
\eqb
  \calF_{i}(\alpha) \Eqn{:=} 1+ \sum_{p=1}^{\infty} \prod_{j=1}^{k}
  { \bigl( b(Q-\alpha, h_{ji}) - b^{2}\bigr)_{p} \over
  \bigl( 1+ b(Q-\alpha, h_{ji}) \bigr)_{p} } \nn \\
  \Eqn{=}
  {}_{k}F_{k-1} \left( 
      \begin{array}{ccccc}
        b(Q-\alpha, h_{1i}) -b^{2}, & \cdots, & -b^{2}, & \cdots, & b(Q-\alpha, h_{ki}) -b^{2} \\
        1+b(Q-\alpha, h_{1i}), & \cdots &
& \cdots, & 1+ b(Q-\alpha, h_{ki})
      \end{array} \bigg\vert \ 1 \,
  \right) \comma \quad 
\eqe
and the $i$-th entry in the lower row is empty. 
We have also defined
\eqb
  (x)_{p} = x(x+1) \cdots (x+p-1) = { \Gamma(x+p) \over \Gamma(x) } \comma
\eqe  
 $\gamma(x) = \Gamma(x)/\Gamma(1-x)$ and
\eqb
   h_{ij} := h_{i} - h_{j} \comma \qquad 
   h_{j} = \omega_{1} - e_{1} - \cdots - e_{j-1} \period
\eqe

We now concentrate on the case of  $k=4$.   
Since $\alpha_0 = -b \omega_2$
for  $p=5, q=7$ of our interest, we first evaluate 
$C^{-b\omega_{2}}_{-be_{0}, -b\omega_{2}}$ with $b$ being generic.
Taking into account the fact that 
the coefficients in front of $\calF_{i}$ in (\ref{defC}) vanish for $i=2,4$, 
we  find after some algebras that
\eqb\label{C3}
  C^{-b\omega_{2}}_{-be_{0}, -b\omega_{2}}
  \Eqn{=} \Bigl( { \pi \tilde{\mu} \over \gamma(-b^{2}) }\Bigr)^{3} 
  { \gamma(1+b^{2}) \over \gamma(2+2b^{2}) }
  \Bigl[  { \gamma(2+3b^{2})\over \gamma(4+5b^{2})} \calF_{1}^{2}(-b\omega_{2})
  +   { \gamma(-2-3b^{2})\over \gamma(-b^{2})} \calF_{3}^{2}(-b\omega_{2})
  \Bigr] \comma \nn \\
   \Eqn{=}  \Bigl( { \pi \tilde{\mu} \over \gamma(-b^{2}) }\Bigr)^{3}
    {\gamma^{2}(1+b^{2}) \over \gamma(2+2b^{2}) \gamma(4+5b^{2})}
    {\Gamma^{2}(3+4b^{2}) \over \Gamma^{2}(2+2b^{2})} \cdot (-2) \cos(2\pi b^{2})
    \comma
\eqe
where 
\eqb
 \calF_{1}(-b\omega_{2}) \Eqn{=}
  {}_{4}F_{3} \biggl( 
      \begin{array}{cccc}
        -b^{2}, & -1-2b^{2}, & -2-4b^{2}, & -3-5b^{2} \\
       & -b^{2}, & -1-3b^{2}, & -2-4b^{2}  
      \end{array} \bigg\vert \ 1 \,
  \biggr)  \nn \\
  \Eqn{=} { \Gamma(-1-3b^{2}) \Gamma(3+4b^{2}) \over
       \Gamma(-b^{2}) \Gamma(2+2b^{2})} \comma  \nn \\ 
   \calF_{3}(-b\omega_{2}) \Eqn{=}
  {}_{4}F_{3} \biggl( 
      \begin{array}{cccc}
        -b^{2}, & 2+2b^{2}, & 1+b^{2}, & -1-2b^{2} \\
       & 3+3b^{2}, & 2+2b^{2}, & -b^{2} 
      \end{array} \bigg\vert \ 1 \,
  \biggr)   \\
  \Eqn{=} { \Gamma(3+3b^{2}) \Gamma(3+4b^{2}) \over
       \Gamma(2+2b^{2}) \Gamma(4+5b^{2})}    \period \nn
\eqe
To obtain the normalized structure constant through (\ref{eq:norm3pt}), 
we next need $B(-be_{0})$, which is evaluated by using
\eqb
   {A(\alpha) \over A(2Q-\alpha)}
   \Eqn{=} \Bigl( \pi \tilde{\mu} \gamma(b^{2}) \Bigr)^{2(\alpha-Q,\rho)/b} \prod_{e>0}
   \frac{1}{b^{2}} { \gamma\bigl(1-b(\alpha-Q,e) \bigr) \over
                             \gamma\bigl(1+b^{-1}(\alpha-Q,e) \bigr)} \comma                    
\eqe
as
\eqb\label{Bbe0}
   B^{2}(-be_{0}) = \Bigl( \pi \tilde{\mu} \gamma(b^{2}) \Bigr)^{-6}
    b^{-24}\Bigl( {3+4b^{2} \over 4+5b^{2}} \Bigr)^{2} 
    \gamma^{-2}(1+b^{2}) \gamma(3+3b^{2}) \gamma(5+5b^{2})
    \period
\eqe
Combining (\ref{C3}) and (\ref{Bbe0}), we find the normalized 
structure constant,
\eqb 
   C^{-b\omega_{2}; \ {\rm normalized}}_{-be_{0}, -b\omega_{2}} 
    \Eqn{=}  B(-be_{0})   C^{-b\omega_{2}}_{-be_{0}, -b\omega_{2}}  \\
   \Eqn{=}  -2 \cos(2\pi b^{2}) \cdot (3+4b^{2})(4+5b^{2})
    {\Gamma^{2}(3+4b^{2}) \over \Gamma^{2}(2+2b^{2})}
    {\gamma(1+b^{2}) \over \gamma(2+2b^{2})}
    \sqrt{ \gamma(3+3b^{2}) \over \gamma(5+5b^{2})}
   \period \nn
\eqe
The factors of $\pi \tilde{\mu} $ have been canceled, as they should. 

Finally, the normalized structure constant $C_{\phi_0  \phi_{(1,1, {\rm adj})}  \phi_0}$
for $p=5, q=7$ and $k=4$
 is obtained by  plugging $b^{2} = -5/7$ in the above expression,
 \eqb\label{eq:SC}
    C_{\phi_0  \phi_{(1,1, {\rm adj})}  \phi_0}
    = -\frac{6}{49} \cos\Bigl( \frac{10}{7} \pi \Bigr) 
    {\Gamma^2(\frac{1}{7}) \gamma(\frac{2}{7}) \over
    \Gamma^2(\frac{4}{7}) \gamma(\frac{4}{7}) }
    \sqrt{ \gamma(\frac{6}{7}) \over \gamma(\frac{10}{7})}
    \approx 1.31083 \sqrt{-1} \period
\eqe
This is purely imaginary since $\gamma(10/7) < 0$.
In subsection \ref{subsection:Fn=7}, 
$\phi_{0}$ and $\phi_{(1,1, {\rm adj})}$ for $WA_3^{(5,7)}$
are denoted by $\hat{\Phi}_{0}$ and $\hat{\Phi}$, respectively.

%
%
\def\thebibliography#1{\list
{[\arabic{enumi}]}{\settowidth\labelwidth{[#1]}\leftmargin\labelwidth
\advance\leftmargin\labelsep
\usecounter{enumi}}
\def\newblock{\hskip .11em plus .33em minus .07em}
\sloppy\clubpenalty4000\widowpenalty4000
\sfcode`\.=1000\relax}
\let\endthebibliography=\endlist
\vspace{3ex}
\begin{center}
{\large\bf References}
\end{center}
\par \vspace*{-2ex}

\end{document}